\documentclass[hyper,11pt,paper]{article}
\usepackage{jheppub}
\usepackage[usenames,dvipsnames,table]{xcolor}
\usepackage{graphicx,amsmath,amssymb,multirow,array,bm,mathrsfs,bbold,bbm}
\usepackage{appendix}
\usepackage{footmisc}
\usepackage{epsf,amsfonts,slashed,soul}
\usepackage{lipsum}
\usepackage{subfig}
\usepackage{pifont}
\usepackage{bbold}
\usepackage{dsfont}
\usepackage{tikz}
\usepackage{xcolor}
\usepackage{verbatim}
\usepackage{placeins}
\allowdisplaybreaks
\newcommand{\bea}{\begin{eqnarray}}
\newcommand{\eea}{\end{eqnarray}}

\newcommand{\be}{\begin{equation}}
\newcommand{\ee}{\end{equation}}
\newcommand{\beq}{\begin{equation}}
\newcommand{\eeq}{\end{equation}}
\newcommand{\ba}{\begin{eqnarray}}
\newcommand{\ea}{\end{eqnarray}}

\title{Changing states in holography: From modular Berry curvature to the bulk symplectic form}

\author[a]{Bart{\l}omiej Czech,}
\author[b]{Jan de Boer,}
\author[a,b]{Ricardo Esp\'indola,}
\author[b]{Bahman Najian,}
\author[b]{Jeremy van der Heijden,}
\author[c]{and Claire Zukowski}
\affiliation[a]{Institute for Advanced Study, Tsinghua University, Beijing 100084, China}
\affiliation[b]{Institute for Theoretical Physics, University of Amsterdam, Science Park 904, Postbus 94485, 1090 GL Amsterdam, The Netherlands}
\affiliation[c]{Department of Physics and Astronomy, University of Minnesota Duluth,
Duluth, MN 55812, USA}

\emailAdd{bartlomiej.czech@gmail.com}
\emailAdd{J.deBoer@uva.nl}
\emailAdd{ricardo.esro1@gmail.com}
\emailAdd{b.najian@uva.nl}
\emailAdd{j.j.vanderheijden2@uva.nl}
\emailAdd{czukowsk@d.umn.edu}

\abstract{We present a new perspective on bulk reconstruction using Berry phases in the boundary CFT. Our parallel transport of modular Hamiltonians is associated to a trajectory in the space of states, which we obtain from the insertion of a source in the Euclidean path integral. Using a modular version of the extrapolate dictionary and the equivalence between modular flow in the boundary and the bulk, we show that the expectation value of the modular Berry curvature on the boundary agrees with an appropriately defined bulk symplectic form associated to the entanglement wedge. In addition, we derive a quantum information metric on the space of density matrices from the Berry curvature, which is related to the canonical energy in the bulk. We also explore the case where a state change reduces to a shape change, uncovering the coadjoint orbit structure of kinematic space in higher dimensions.}

\begin{document}
\maketitle

\section{Introduction}

The growing interface between quantum information theory and gravity has shed new light on many aspects of quantum gravity; for recent reviews see \cite{Iqbal:2016qyz, Rangamani:2016dms,Harlow:2018fse,Headrick:2019eth,Chen:2021lnq}. 
Within the realm of holography, it has been fruitful to search for bulk duals of quantum information theoretic concepts on the boundary, so as to add new entries to the AdS/CFT dictionary. Contrasted with earlier results in AdS/CFT, the quantum information theory-based part of the dictionary often has a more direct connection to bulk geometry.

In this paper, we continue the approach of deriving new AdS/CFT dictionary entries from quantum information theoretic quantities on the boundary side. We will investigate a particular new quantum information theoretic boundary quantity, which adapts the Berry parallel transport \cite{Berry:1984jv} to trajectories in the space of global states. Unlike Berry transport for pure states in quantum mechanics, this parallel transport transforms operators associated to a spatial subregion. This process has been dubbed \emph{modular} Berry transport because it relies on entanglement properties of subregions, specifically on how the modular Hamiltonian is glued together across different choices of subregion. 

Modular Berry transport has been studied in some detail for trajectories defined over kinematic space \cite{Czech:2015qta}---ones where boundary subregions vary in shape or location~\cite{Czech:2017zfq, Czech:2019vih}. In this case there is a direct bulk geometric dual: The Berry phase reproduces lengths of bulk curves that can be reached by extremal surfaces, and the Berry curvature is related to a bulk curvature. A close cousin of the modular parallel transport generator was recently shown to act in three-dimensional bulk geometries as the generator of ordinary parallel transport, which is described by General Relativity \cite{Chen:2022fbg}.

Our setting here is different from those earlier works. We consider modular parallel transport along trajectories, which visit varying global \emph{states}  rather than varying locations or shapes of boundary subregions. This approach was initiated in a more restricted setting in \cite{deBoer:2021zlm}. There, we showed that the curvature associated to a particular state-changing modular Berry transport could be identified with an appropriately defined symplectic form associated to an entanglement wedge. In that case, state deformations were implemented through the action of a large diffeomorphism, whose form was dictated by the Virasoro symmetry of a CFT$_2$. (Berry phases on the Virasoro algebra were likewise considered in \cite{Oblak:2017ect, Akal:2019hxa, Das:2020goe, Banerjee:2022jnv}.) This setting further revealed a connection to an auxiliary symplectic geometry derived from the group theory of the Virasoro algebra: a coadjoint orbit. The triality between the Berry curvature, the entanglement wedge symplectic form, and the Kirillov-Kostant symplectic form on an appropriate orbit (see also~\cite{Chagnet:2021uvi} for a similar triality for bulk duals of complexity) revealed an interplay between group theory and quantum information in this case, giving an additional handle on an important bulk geometric quantity of interest.

Our aim in this paper is to set up Berry transport for a broad class of state deformations in any dimension. Based on previous results in two dimensions, one might imagine this to be a straightforward task. However, the power of group theory to describe certain state-changing transformations in two dimensions also presents a limitation in its generalization. To generalize state-changing Berry transport to a larger class of state changes including state changes in higher dimensions, one must invoke a very different toolkit. In the present work we make use of the Euclidean path integral to implement state changes (analogously to~\cite{Belin:2018fxe, Belin:2018bpg} in the case of pure states, or see~\cite{Kirklin:2019ror} for a different version of parallel transport based on the Uhlmann phase\footnote{Excellent summaries of the different types of transport in quantum mechanical state spaces, including Berry and Uhlmann transport, are given in \cite{phasesbook} and \cite{andersson}. Note, however, that those works assume that the Hilbert space is finite-dimensional. Infinite-dimensional Hilbert spaces can give rise to subtleties, see for example \cite{deBoer:2021zlm}.}). We also use some new (from the perspective of modular Berry transport) techniques such as modular Fourier decompositions, the KMS condition from modular theory, as well as (from the bulk side) the equivalence between bulk and boundary modular flow and the modular extrapolate dictionary. These tools have been useful in proving the ANEC and the quantum null energy condition~\cite{Faulkner:2016mzt, Balakrishnan:2017bjg} and in setting the stage for a modular approach to bulk recontruction~\cite{Faulkner:2017vdd, Balakrishnan:2017bjg, Chen:2018rgz}. Intriguingly, though we employ very different techniques from group theory, coadjoint orbits and Chern-Simons theory as utilized in~\cite{deBoer:2021zlm}, the end result is similar: The expectation value of the Berry curvature in the global pure state is equal to the symplectic form associated to an entanglement wedge.

\begin{figure}[t!]
\centering
\includegraphics[width=0.8\columnwidth]{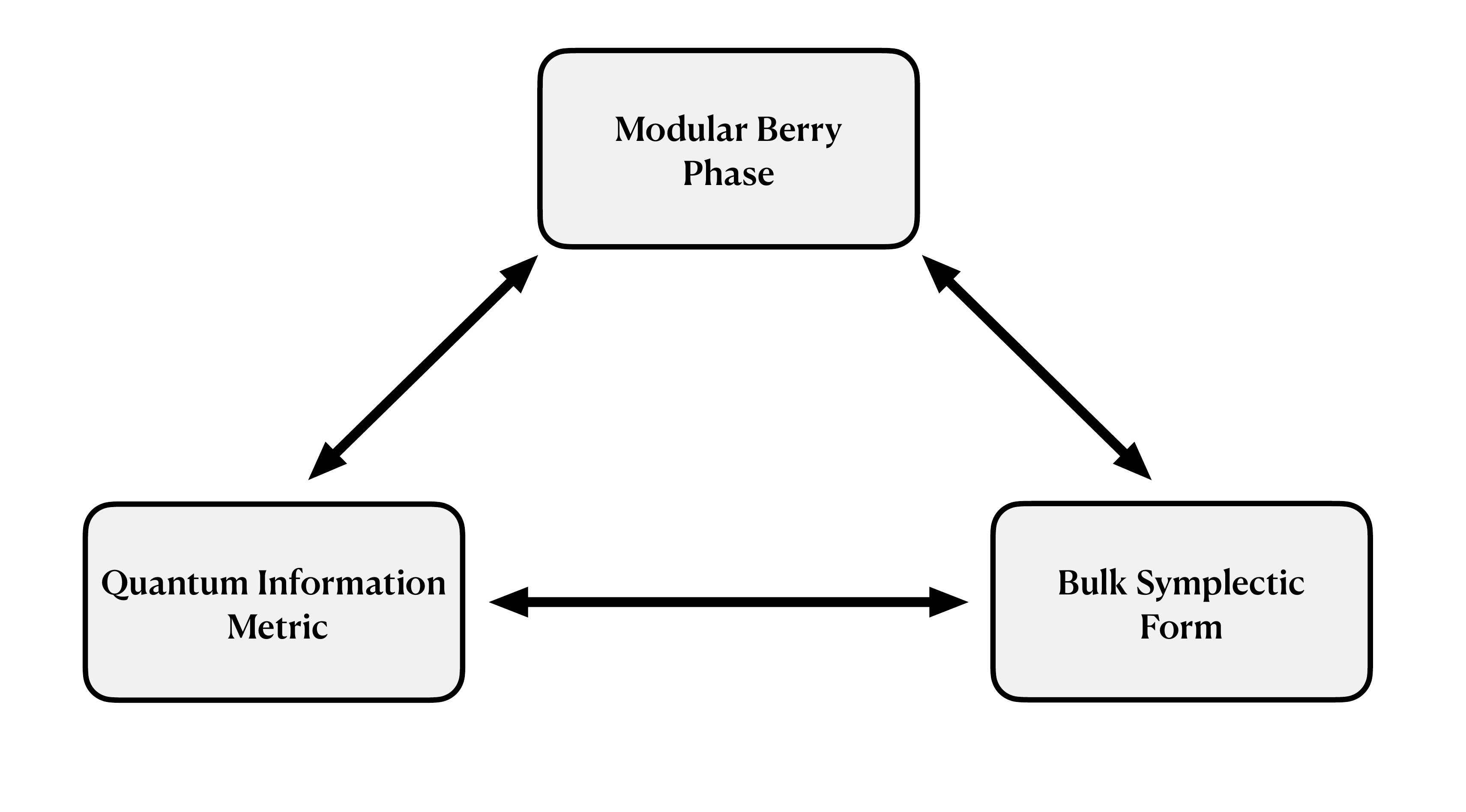}
   \caption{Modular Berry transport provides a framework that encodes information about not only the bulk symplectic form, but also the quantum information metric.} 
    \label{fig:Triangle}
\end{figure}

Using a similar framework, we can also extract from the full Berry curvature a symmetric quantity. We show that on the boundary, this describes a metric on the space of density matrices, often referred to as the quantum Fisher information metric (this also goes by other names). In the bulk, we extract this from the bulk symplectic form by taking a Lie derivative with respect to the generator of modular flow. This describes the canonical energy, which has been used as a tool for deriving the bulk equations of motion from entanglement entropy~\cite{Faulkner:2013ica, Lashkari:2015hha,Faulkner:2017tkh}. In the end, we see that the modular Berry phase incorporates more information beyond simply the bulk symplectic form, as is represented in the triangle in Figure~\ref{fig:Triangle}.

Along the way, we can make contact with Berry transport in the shape-changing case, now generalized to higher dimensions. We do so in two ways: first, by considering the specific case of state deformations sourced by the stress tensor, which incorporates shape changes. Next, we act with symmetry generators of the higher dimensional conformal algebra, in a direct generalization of the techniques of~\cite{deBoer:2021zlm}. In doing so, we relate the Berry curvature for the higher-dimensional shape-changing case to the Kirillov-Kostant symplectic form on a coadjoint orbit. The full non-abelian Berry curvature lives on the coset space that is relevant for the higher dimensional version of kinematic space, the space of causal diamonds in a CFT~\cite{Czech:2014ppa, Czech:2015qta, Czech:2016xec, deBoer:2015kda, deBoer:2016pqk, Huang:2020cye, Huang:2021qkm}. The connection to the Kirillov-Kostant symplectic form relies on the fact that in this case, unlike for general state transformations, the deformations which implement parallel transport lie in the symmetry algebra of the boundary.\\

\noindent {\bf Outline}: We set the stage in Section~\ref{sec:coherentstate} by reviewing modular Berry transport and state preparation using the Euclidean path integral. After introducing some of the language of modular flow and modular Fourier decomposition, we use these tools to derive the modular Berry curvature for general state deformations. We also introduce a symmetric derivative of the Berry curvature (see Appendix~\ref{sec:relativeentropy} for the quantum information theoretic interpretation). Next, we extend these quantities into the bulk in Section~\ref{sec:bulksymp} using the modular extrapolate dictionary. We show explicitly for operators sourcing bulk scalar fields that this computes the bulk symplectic form (see Section~\ref{sec:stresstensordef} for a generalization beyond the scalar case). The symmetric offshoot is related to the bulk canonical energy. Section~\ref{sec:explicitex} presents some explicit examples of the general formalism of the previous sections. Specifically, we consider in Section~\ref{sec:stresstensordef} the case of a stress tensor source, which in general implements a change of metric but also includes the shape-changing case. An explicit bulk computation of the symplectic form for the shape-changing sub-case is given in Appendix~\ref{sec:gravsympshape}. Finally, in Section~\ref{sec:coadjoint} we explicitly consider the higher-dimensional shape-changing case by acting with symmetry generators, and elucidate the connection to coadjoint orbits. Our conventions for the conformal algebra are presented in Appendix~\ref{appendixA}.
\section{Berry curvature for coherent state deformations} \label{sec:coherentstate}
First we consider a parallel transport problem purely defined on the boundary. In Section~\ref{sec:transport}, we review modular Berry transport. This parallel transport problem concerns modular Hamiltonians, which undergo deformations. We will subsequently apply this formalism to deformations that change the global state on the boundary in arbitrary dimension. In Section~\ref{sec:EuclPI}, we review how to construct such state deformations using coherent states and the Euclidean path integral, and in Sections~\ref{sec:BerryCurv} and~\ref{sec:QImetric} we derive new results for the modular Berry curvature and quantum information metric for state deformations. Our results make convenient use of modular eigenstates and a modular Fourier basis.

\subsection{Modular Berry transport}\label{sec:transport}

We begin by reviewing modular Berry transport, which is the starting point for much of our analysis~\cite{Czech:2017zfq, Czech:2019vih}. Consider a subregion $A$ along a time-slice of a $d$-dimensional CFT. Let $|\Psi\rangle$ be a pure state defined on the whole space. We can obtain a reduced density matrix associated to $A$ by tracing over the complement $\bar{A}$:
\beq \label{eq:densitymatrix}
\rho_A =\mathrm{tr}_{\bar{A}}|\Psi\rangle \langle \Psi |~.
\eeq
Since in this paper we will only be concerned with mixed states, we henceforth drop the subscript $A$. Operator $\rho$ should be understood as the reduced density operator associated with a subregion, rather than a pure state. 

We can define from this a modular Hamiltonian $H_{\rm mod}$, which is related to the density matrix through
\beq 
\rho = e^{-H_{\rm mod}}~.
\eeq
The modular Hamiltonian, like the reduced density matrix, depends on the specified subregion $A$. More generally, there is an algebra of observables $\mathcal{A}$ associated to the region $A$. There is also an algebra $\mathcal{A}_0$ of `modular zero modes' $Q_i$ which commute with the modular Hamiltonian,
\be [Q_i, H_{\rm mod}] = 0~.\ee
Given an operator $\mathcal{O}(0)$ associated to the region, the flow by a modular zero mode 
\be \mathcal{O}(s_i) = V^\dagger(s_i) \mathcal{O}(0) V(s_i)~, \indent V(s_i)=e^{-i s_i Q_i}~, \ee
will leave the expectation value of the operator unchanged. It will also map the operator to another operator in the same region, thus leaving the algebra unchanged.

Now, consider a family of modular Hamiltonians $H_{\rm mod}(\eta)$ that depends on some parameter $\eta$. For instance, $\eta$ could specify the shape or location of the subregion $A$ as it is slowly varied~\cite{Czech:2017zfq, Czech:2019vih, Chen:2022nwf}. Another possibility is to fix $A$ but generate a family of modular Hamiltonians by varying the global state $|\Psi\rangle$, as was considered in two dimensions in~\cite{deBoer:2021zlm}. We can consider a ``modular'' parallel transport process by studying the transport of an operator associated to the subregion as $\eta$ is varied. Note that ultimately, the modular Hamiltonian is not well-defined in QFT. In this paper, we will implicitly introduce a small cutoff to have a well-defined notion of modular Hamiltonian. 

Consider diagonalizing the modular Hamiltonian in a given basis, whose details are unimportant for the rest of the computation:
\be H_{\rm mod} = U^\dagger \Delta U~, \label{eq:Hmoddiag}\ee
where the spectral piece $\Delta$ is a diagonal matrix, and $U$ implements a change of basis. Both $U$ and $\Delta$ will in general depend on $\eta$. Taking the derivative with respect to $\eta$ gives
\be \label{eq:dotH} \dot{H}_{\rm mod} = [\dot{U}^\dagger U, H_{\rm mod}]+ U^\dagger \dot{\Delta} U~.\ee
Here, the dot is a derivative with respect to $\eta$. Notably, this equation is invariant under the action of a modular zero mode flow. One can view the modular zero mode action as a redundancy along the path that an operator is being transported through. This is analogous to the ordinary Berry phase redundancy (for the pure state case) in quantum mechanics. Under a finite displacement in the parameter space that is closed (i.e., one where $\eta$ is the same at the initial and end point), an operator may not return exactly to itself, but rather to itself up to a flow by a modular zero mode.

Define a projection $P_0$ that sends an operator in the algebra of observables to its zero mode component:
\beq
P_0:\mathcal{A}\to \mathcal{A}_{0}~.
\eeq
It is straightforward to construct such a projection operator in finite-dimensional settings, which is canonical given some choice of inner product for which $H_{\rm mod}$ is Hermitian. Subtleties concerning uniqueness of this projection operator in two dimensions---where the observable algebra is infinite-dimensional---were discussed in \cite{deBoer:2021zlm}. $\dot{U}^\dagger U$ is, up to an additive zero mode, the generator of parallel transport. Indeed, the zero mode projection $P_0(\dot{U}^\dagger U)$ transforms as a gauge field, so we can think of $P_0(\dot{U}^\dagger U)$ as a component of the Berry connection~\cite{Czech:2019vih}. Said differently, the modular parallel transport condition reads:
\beq 
P_0(\dot{U}^\dagger U)=0~.
\eeq

We will find it useful to rewrite \eqref{eq:dotH} in the form
\be
\delta H_{\rm mod}-P_0(\delta H_{\rm mod}) =  [X,H_{\rm mod}]~.
\ee
This uses $X$ to represent $\dot{U}^\dagger U$, which---we emphasize again---is the generator of parallel transport plus possibly some zero modes. (It is the parallel transport generator if $P_0(X) = 0$.) $\delta H_{\rm mod}$ is understood to denote $\dot{H}_{\rm mod}$. We also rely on having the image and kernel of the adjoint action $[\cdot, H_{\rm mod}]$ with respect to the modular Hamiltonian be disjoint. This is always true, for instance, in the finite dimensional case of the conformal group that we will consider in Section~\ref{sec:coadjoint}. This implies that $[X,H_{\rm mod}]$ contains no zero modes so that $P_0(\delta H_{\rm mod}) = U^\dagger \dot{\Delta} U$.

Now consider two deformations in different directions in parameter space, which can be generated by operators $X_1$ and $X_2$. We assume these $X_{1,2}$ generate parallel transport in their respective directions, that is $P_0(X_1) = P_0(X_2) = 0$. The holonomy around an infinitesimal loop in parameter space is the Berry curvature (see Appendix B of~\cite{deBoer:2021zlm}):
\be \label{eq:Berryformula}
F=P_0([X_1,X_2])~.
\ee
We will compute this quantity explicitly for state-changing transformations, with the aim of finding an appropriate bulk dual.

\subsection{Coherent state deformations} \label{sec:EuclPI}
We would like to consider a modular Berry setup where the variation of $\eta$ denotes a change of \emph{state} rather than a change of shape or location of the subregion. For a CFT in two dimensions, one particular class of deformations $\delta H_{\rm mod}$ that implement state changes involve elements of the infinite-dimensional Virasoro symmetry algebra. (Due to certain subtleties, it is necessary to employ a continuous version of the Virasoro algebra, which is described by certain non-smooth vector fields on the circle~\cite{deBoer:2021zlm}.) For a CFT in $d>2$ dimensions the story will necessarily be different, since such state-changing transformations no longer lie in the symmetry algebra $\mathfrak{so}(d,2)$. 

To generalize the state-changing Berry construction to accommodate higher-dimensional setups, it will be useful to introduce the language of Euclidean path integrals.\footnote{The Euclidean path integral is also useful for defining a CFT Berry transport process for \emph{pure} states, without restricting to a subregion~\cite{Belin:2018fxe, Belin:2018bpg}. 
One goal of our work is to explicitly adapt this to modular transport for mixed states. For a formal argument involving the mixed state case and a different variety of parallel transport, see~\cite{Kirklin:2019ror}.} 
Specifically, we assume that the state $|\Psi\rangle$ is a coherent state, in the sense that it is prepared by the Euclidean path integral with a background source $\lambda$. We consider deforming the state through the insertion of some operator $\mathcal{O}$ in the path integral:
\be \label{eq:Sdeform}
\delta S = \int d^dx \,\delta \lambda (x) \mathcal{O}(x)~.
\ee
The source $\delta \lambda(x)$ determines the strength of the perturbation. At this point, there is no need to restrict the support of the source, we take it to be anywhere in the Euclidean half-plane. 

\begin{figure}[t!]
\centering
\includegraphics[width=0.6\columnwidth]{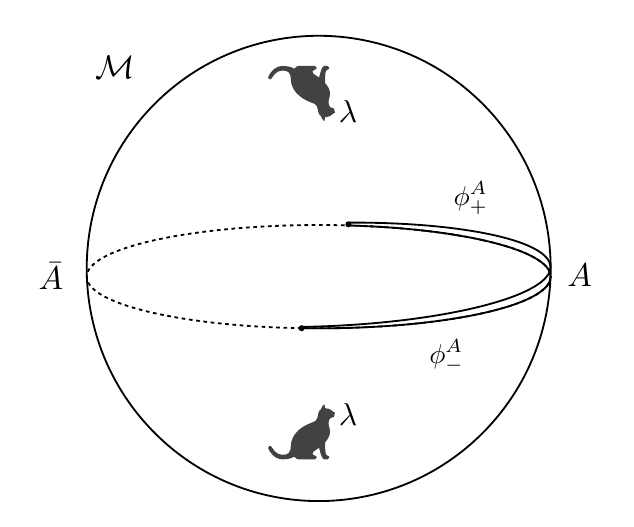}
   \caption{The Euclidean manifold $\mathcal{M}$ that is integrated over to prepare the matrix elements $\langle \phi_+^A | \rho |\phi_-^A \rangle$ of the density matrix. The two hemispheres are glued along the complement region $\bar{A}$, while the boundary conditions at the region $A$ are left open. One can prepare non-trivial coherent states by introducing non-trivial background sources $\lambda$ (which are represented by cats in the figure).}
    \label{fig:pathintegral}
\end{figure}

The perturbation \eqref{eq:Sdeform} leads to a change of the density matrix, and hence of the modular Hamiltonian~\cite{Faulkner:2016mzt}. 
Denoting the collective field content of the theory by $\phi$, one can compute matrix elements of the density matrix \eqref{eq:densitymatrix} by gluing the upper and lower Euclidean half plane along the complement $\bar{A}$ at $t_{E}=0$:
\beq 
\langle \phi_+^A | \rho |\phi_-^A \rangle =\frac{1}{Z}\int_{\phi(0^-)=\phi_-^A}^{\phi(0^{+})=\phi_+^A}[D\phi]\, e^{-S[\phi]}~, \hspace{10pt} \mathrm{where} \hspace{10pt} Z\equiv \int [D\phi]\, e^{-S[\phi]}~.
\eeq
Here, $\phi^{A}_+$ and  $\phi^{A}_-$ denote the value of the field $\phi$ just above and below the subregion $A$ respectively, and $S[\phi]$ is the Euclidean action of the theory with source $\lambda$. One therefore integrates over the full Euclidean manifold $\mathcal{M}$ with a branch cut at the location of the subregion (see Figure \ref{fig:pathintegral}). 

We now perturb the state according to \eqref{eq:Sdeform}. The new density matrix $\rho'$ is given by
\beq \label{eq:densitymatrixper}
\langle \phi_+^A | \rho' |\phi_-^A \rangle =\frac{1}{(Z+\delta Z)}\int_{\phi(0^-)=\phi_-^A}^{\phi(0^{+})=\phi_+^A}[D\phi]\, e^{-S[\phi]-\int d^dx \,\delta \lambda(x) \mathcal{O}(x)}~.
\eeq
Using the geometric series relation 
\beq 
\frac{1}{(Z+\delta Z)} = \frac{1}{Z}\left(1-\frac{\delta Z}{Z}+\ldots\right)~, 
\eeq
and expanding the exponential in \eqref{eq:densitymatrixper}, we find that the change $\delta \rho \equiv \rho'-\rho$ is given by 
\begin{align}
\langle \phi_+^A | & \delta \rho |\phi_-^A \rangle=-\frac{1}{Z}\int_{\phi(0^-)=\phi_-^A}^{\phi(0^{+})=\phi_+^A}[D\phi]\, e^{-S[\phi]}\int d^dx\, \delta \lambda(x):\mathcal{O}(x) :+\ldots ~,
\end{align}
where we have introduced the renormalized operator $:\mathcal{O}:\equiv \mathcal{O}-\langle \mathcal{O}\rangle$. From now on we will omit the notation $:\cdot:$, and assume that all operators are background-subtracted. Hence, up to first order in the source the density matrix changes as
\beq \label{eq:densitymatrixchange}
\delta \rho =-\int d^dx\, \rho \, \delta \lambda(x)\mathcal{O}(x)~.
\eeq

Recall that the modular Hamiltonian $H_{\rm mod}$ is related to $\rho$ by
\beq 
\rho = e^{-H_{\rm mod}}~.
\eeq
Using the integral representation of the logarithm,
\beq 
H_{\rm mod}=-\log \rho = \int_0^{\infty} d\beta \left(\frac{1}{\rho+\beta}-\frac{1}{1+\beta}\right)~,
\eeq
it follows from \eqref{eq:densitymatrixchange} that
\begin{align} \label{eq:deltaHdeltarho}
\delta H_{\rm mod}
&=
- \int_{0}^{\infty} d\beta \left(\frac{1}{\rho+\beta}\delta \rho \frac{1}{\rho+\beta}\right)
=
\int d^dx\, \delta \lambda(x)\int_{0}^{\infty} d\beta  \left(\frac{\rho}{\rho+\beta}\mathcal{O}(x)\frac{1}{\rho+\beta}\right)~.
\end{align}
To proceed, it is useful to use a spectral representation for the density matrix $\rho$. We consider modular frequency states $|\omega\rangle$, which are eigenstates of the modular Hamiltonian:\footnote{As the existence of such states is only guaranteed in type I von Neumann algebras, our analysis presumes that the more realistic settings of type II algebras (semi-classical gravity) and/or type III algebras (quantum field theory) do not alter the overall picture.}
\be \label{eq:eigenstates}
H_{\rm mod}|\omega\rangle=\omega|\omega\rangle~.
\ee
When evaluated in this basis, the change in the modular Hamiltonian takes a relatively simple form. Inserting a resolution of the identity, one finds
\begin{align} \label{eq:deltaHmod1}
\langle \omega | & \delta H_{\rm mod} |\omega' \rangle
=\int d^dx\, \delta \lambda(x) \langle \omega|\mathcal{O}(x)|\omega'\rangle\int_{0}^{\infty} d\beta \left(\frac{e^{-\omega}}{e^{-\omega }+\beta}\frac{1}{e^{-\omega'}+\beta}\right)~.
\end{align}
The integral over $\beta$ can be performed easily. Indeed, we find that  
\beq 
\int_{0}^{\infty} d\beta \left(\frac{e^{-\omega}}{e^{-\omega }+\beta}\frac{1}{e^{-\omega'}+\beta}\right)= \frac{\omega -\omega'}{e^{\omega-\omega'}-1}~.
\eeq
Plugging this back into \eqref{eq:deltaHmod1}, it follows that
\begin{align} \label{eq:deltaHmod2}
\langle \omega | \delta H_{\rm mod} |\omega' \rangle
=\int d^dx\, \delta \lambda(x) n(\omega-\omega') (\omega-\omega')\langle \omega|\mathcal{O}(x)|\omega'\rangle ~,
\end{align}
where here we have introduced the quantity
\be \label{eq:Boltzmann}
n(\omega)\equiv \frac{1}{e^{\omega}-1}~,
\ee
which will be convenient later.
This gives a relatively simple expression for the change in modular Hamiltonian in terms of the matrix elements of the operator $\mathcal{O}$. 

Recall that the modular parallel transport problem relies on defining projection $P_0$ that sends an operator to its zero mode component. There is an ambiguity in how to define this projection. A natural choice is to take the diagonal matrix elements of the operator and multiply by the eigenstate $|\omega\rangle \langle \omega|$:\footnote{ Under general circumstances, the integral in (\ref{eq:projectionop}) might involve a non-trivial density of states. The attendant degeneracies among states $|\omega\rangle$ generically arise from additional symmetries, which commute with $H_{\rm mod}$. If so, one can extend $H_{\rm mod}$ to a complete set of commuting operators and declare $\omega$ to denote the corresponding complete set of quantum numbers. In this paper we assume that any degeneracies in the spectrum have been accounted for in this fashion, and do not include explicit factors of the density of states.
}
\beq \label{eq:projectionop}
P_0(\mathcal{O})\equiv\int d\omega \, \langle \omega | \mathcal{O} |\omega \rangle |\omega \rangle \langle \omega |~.
\eeq
This procedure defines a diagonal operator, which commutes with the modular Hamiltonian since $H_{\rm mod}$ is diagonal in its own eigenbasis. It is easy to check that the projection \eqref{eq:projectionop} satisfies
\beq \label{eq:projprop}
P_0(H_{\rm mod})=H_{\rm mod}~, \hspace{10pt} P_0([H_{\rm mod},X])=0~.
\eeq
In other words, it is indeed the case that the kernel and image of the adjoint action $[\cdot, H_{\rm mod}]$ with respect to the modular Hamiltonian are disjoint.

We can now use $P_0$ to define the parallel transport problem. We first subtract off the zero mode part of $\delta H_{\rm mod}$, which is given by the diagonal component of \eqref{eq:deltaHmod2}:
\beq \label{eq:deltaHzero-mode}
P_0(\delta H_{\rm mod}) =\int d^dx\, \delta \lambda(x) P_0(\mathcal{O}(x))~.
\eeq
Using the fact that
\be \label{eq:commrelation}
\langle \omega | [X,H_{\rm mod}]|\omega' \rangle = (\omega'-\omega)\langle \omega |X|\omega' \rangle~,
\ee
we recognize that the factor $\omega-\omega'$ in \eqref{eq:deltaHmod2} comes from a commutator. Indeed, we can choose an $X$ with matrix elements
\be \label{eq:Xoperator}
\langle \omega | X |\omega' \rangle
=- \int d^dx\, \delta \lambda(x) n(\omega-\omega') \langle \omega|\mathcal{O}(x)|\omega'\rangle~.
\ee
We additionally assume that $X$ is zero mode free, $P_0(X)=0$, which also implies $P_0(\delta H_{\rm mod})=0$ by~\eqref{eq:deltaHzero-mode}. Then, by comparing~\eqref{eq:commrelation} and~\eqref{eq:Xoperator} with \eqref{eq:deltaHmod2}, we see that $X$ satisfies the transport equation
\beq \label{paralleltransport}
\langle \omega | (\delta H_{\rm mod}-P_0(\delta H_{\rm mod}))|\omega' \rangle = \langle \omega | [X,H_{\rm mod}]|\omega' \rangle~.
\eeq
Since $X$ is assumed to be zero mode free, we can identify it with the generator of parallel transport whose commutators compute the modular Berry curvature \eqref{eq:Berryformula}. 

\subsection{Berry curvature}\label{sec:BerryCurv}
Now that we have computed the matrix elements of $\delta H_{\rm mod}$ in the modular eigenstate basis and derived the generator $X$ of parallel transport, we would like to compute from this the Berry curvature. Recall that given two infinitesimal deformations $\delta_1 \lambda, \delta_2 \lambda$ and corresponding zero-mode free parallel transport generators $X_1, X_2$, the Berry curvature is given by
\be \label{eq:Berryformula2}
F=P_0([X_1,X_2])~.
\ee

To further evaluate this expression it is useful to decompose the operator $\mathcal{O}$ in a `modular Fourier basis,' where the action of the modular Hamiltonian is simple. Such a basis was previously used in the context of bulk reconstruction in \cite{Faulkner:2017vdd}. Let us first consider the modular flow associated to the algebra $\mathcal{A}$ and state $|\Psi\rangle$, defined by the operation
\beq \label{eq:modularflow}
\mathcal{O}\in \mathcal{A} \to \mathcal{O}_s = e^{iH_{\rm mod}s}\mathcal{O} e^{-iH_{\rm mod}s}\in \mathcal{A}~.
\eeq
One can use the modular flow to make a Fourier decomposition of the form 
\beq \label{eq:frequencyoperators}
\mathcal{O}_{\omega} = \int_{-\infty}^{\infty}ds \, e^{-i\omega s}\mathcal{O}_s~,
\eeq
where the operators $\mathcal{O}_{\omega}$ are labeled by some modular frequency $\omega$. We can now decompose an operator $\mathcal{O}$ in terms of the modular Fourier basis as
\beq \label{eq:Fourierdecomposition}
\mathcal{O}=\frac{1}{2\pi}\int d\omega\, \mathcal{O}_{\omega}~.
\eeq
Note that the operators $\mathcal{O}_{\omega}$ should always be viewed as being integrated against some suitable function of the frequency $\omega$ to get finite expectation values. Therefore, using the modular Fourier basis directly will introduce some intermediate $\delta$-functions\footnote{Note that the Fourier zero mode $\mathcal{O}_0$ commutes with the modular Hamiltonian, but it is not the same as applying the zero mode projection $P_0(\mathcal{O})$. They differ by an infinite normalization factor coming from the extra $\delta$-function:
\beq
\mathcal{O}_0=P_0(\mathcal{O}_0)=2\pi \delta(0) P_0(\mathcal{O})~,
\eeq
which reflects the fact that $\mathcal{O}_0$ by itself is in some sense a singular operator.}  in the computation, but the final answer for the curvature will be finite.

The action of modular flow \eqref{eq:modularflow} on $\mathcal{O}_\omega$ is particularly simple. By shifting the integration variable in \eqref{eq:frequencyoperators} we find that 
\beq \label{eq:modularevolution}
e^{iH_{\rm mod}t} \mathcal{O}_{\omega} e^{-iH_{\rm mod}t} = e^{i\omega t} \mathcal{O}_{\omega}~.
\eeq
Plugging this into the formula for the commutator
\beq 
[H_{\rm mod}, \mathcal{O}_{\omega}] = -i \frac{d}{dt}\Big|_{t=0}  e^{iH_{\rm mod}t} \mathcal{O}_{\omega} e^{-iH_{\rm mod}t}~,
\eeq
gives the relation 
\beq \label{eq:eigenequation}
[H_{\rm mod}, \mathcal{O}_{\omega}]=\omega \mathcal{O}_{\omega}~.
\eeq
We conclude that the operators \eqref{eq:frequencyoperators} constitute a formal spectral decomposition of the adjoint action of $H_{\rm mod}$. 

The matrix elements of $\mathcal{O}_{\omega}$ in the modular frequency basis obey
\begin{align}
\langle \omega'| \mathcal{O}_{\omega} | \omega''\rangle &= \int_{-\infty}^{\infty} ds \, e^{i(\omega' -\omega -\omega'')s} \langle \omega'|\mathcal{O} | \omega''\rangle= 2\pi \delta(\omega'-\omega-\omega'')  \langle \omega'|\mathcal{O} | \omega''\rangle~, \label{eq:frequencies}
\end{align}
so they are only non-zero when the frequencies satisfy the condition $\omega =\omega'-\omega''$. This can be used to our advantage. In particular, one can use \eqref{eq:frequencies} to show that
\beq \label{eq:indentity}
\int d\omega'' f(\omega'') \langle \omega'| \mathcal{O}_{\omega} | \omega''\rangle = 2\pi f(\omega'-\omega) \langle \omega'|\mathcal{O} | \omega'-\omega\rangle=  \int d\omega'' f(\omega'-\omega'') \langle \omega'| \mathcal{O}_{\omega''} | \omega'-\omega\rangle~,
\eeq
and similarly 
\beq
\int d\omega'' f(\omega'') \langle \omega''| \mathcal{O}_{-\omega} | \omega'\rangle =  \int d\omega'' f(\omega'-\omega'') \langle \omega'-\omega| \mathcal{O}_{-\omega''} | \omega'\rangle~
\eeq
for any function $f=f(\omega)$. 

This identity can be used to transform an integral over modular frequency states to an integral over modular frequency operators. Let us first decompose the operator $X_1,X_2$ into modular Fourier modes $X_{1,\omega_1},X_{2,\omega_2}$, and compute the commutator
\begin{align} \label{eq:[X1,X2]}
\langle \omega| &[X_{1,\omega_1},X_{2,\omega_2}] |\omega' \rangle=\int d^dx \int d^dx'\, \delta_1 \lambda(x) \delta_2 \lambda(x') \int d\omega'' \langle \omega|\mathcal{O}_{\omega_1}(x)|\omega''\rangle \langle \omega''| \mathcal{O}_{\omega_2}(x')|\omega'\rangle \nonumber \\
& \hspace{20pt} \times n(\omega-\omega'') n(\omega''-\omega') - (1\leftrightarrow 2)~.
\end{align}
We have inserted a complete basis of states and used the expression \eqref{eq:Xoperator}. We will now consider the diagonal part of \eqref{eq:[X1,X2]}. First note that from \eqref{eq:frequencies}, 
$\langle \omega|\mathcal{O}_{\omega_1}(x)|\omega''\rangle \langle \omega''| \mathcal{O}_{\omega_2}(x')|\omega\rangle$ is proportional to $\delta(\omega-\omega_1-\omega'') \delta(\omega''-\omega_2-\omega)$. Thus, it is only non-zero when $\omega_1=-\omega_2$. We are therefore allowed to multiply the equation with a term $\delta(\omega_1+\omega_2)\delta(0)^{-1}$. The extra insertion of $\delta(0)$ will cancel at the end of the computation, when we write the answer in terms of the original operators. Using the identity \eqref{eq:indentity} we find that 
\begin{align} 
\int & d\omega''n(\omega-\omega'') n(\omega''-\omega) \langle \omega|\mathcal{O}_{\omega_1}(x)|\omega''\rangle \langle \omega''| \mathcal{O}_{\omega_2}(x')|\omega\rangle \nonumber \\
&= \delta(\omega_1+\omega_2) \delta(0)^{-1} \int d\omega'' n(-\omega'') n(\omega'') \langle \omega|\mathcal{O}_{\omega''}(x)|\omega-\omega_1\rangle \langle \omega-\omega_1| \mathcal{O}_{-\omega''}(x')|\omega\rangle~.
\end{align}
By integrating over the modular frequencies $\omega_1,\omega_2$ on both sides of the equality using \eqref{eq:Fourierdecomposition}, and then removing a resolution of the identity, one obtains
\begin{align} 
\int  d\omega'' n(\omega-\omega'') &n(\omega''-\omega) \langle \omega|\mathcal{O}(x)|\omega''\rangle  \langle \omega''| \mathcal{O}(x')|\omega\rangle \nonumber \\
&= \mathcal{N}^{-1} \int d\omega'' n(-\omega'') n(\omega'') \langle \omega|\mathcal{O}_{\omega''}(x) \mathcal{O}_{-\omega''}(x')|\omega\rangle~,
\end{align}
where $\mathcal{N}\equiv (2\pi)^2 \delta(0)$. Putting this back into the expression for the commutator $[X_1,X_2]$, one finds
\begin{align}
\langle & \omega| [X_{1},X_{2}] |\omega \rangle \nonumber \\
&=\mathcal{N}^{-1}\int d^dx \int d^dx'\, \delta_1 \lambda(x) \delta_2 \lambda(x') \int d\omega'' n(-\omega'') n(\omega'') \langle \omega|[\mathcal{O}_{\omega''}(x), \mathcal{O}_{-\omega''}(x')]|\omega\rangle~.
\end{align}
Since the operator $[\mathcal{O}_{\omega''}(x), \mathcal{O}_{-\omega''}(x')]$ is diagonal already, the projection operator $P_0$ leaves it invariant. We conclude that the Berry curvature \eqref{eq:Berryformula}  is given by
\begin{align} \label{eq:F}
F=\mathcal{N}^{-1}\int d^dx \int d^dx'\, \delta_1 \lambda(x) \delta_2 \lambda(x') \int d\omega \, n(-\omega) n(\omega)[\mathcal{O}_{\omega}(x),\mathcal{O}_{-\omega}(x')]~.
\end{align}
This formula is one of the main results of this section, and it provides a useful representation of the curvature associated to coherent state deformations of the form \eqref{eq:Sdeform}. 
 
Note that this modular Berry curvature $F$ is operator-valued, due to the fact that our transport problem is suited to density matrices, instead of pure states. In fact it is easy to verify that the curvature is a zero mode, i.e., $F\in \mathcal{A}_{0}$. By virtue of the Jacobi identity together with~\eqref{eq:eigenequation},
\beq 
[H_{\rm mod}, [\mathcal{O}_{\omega},\mathcal{O}_{-\omega}] ] = [\mathcal{O}_{\omega},[H_{\rm mod},\mathcal{O}_{-\omega}] ]-[\mathcal{O}_{-\omega},[H_{\rm mod},\mathcal{O}_{\omega}] ] =0~,
\eeq
which shows that the curvature indeed satisfies $[H_{\rm mod},F]=0$. Moreover, the expression \eqref{eq:F} is anti-symmetric under interchanging $1$ with $2$. This can be most easily seen by substituting $\omega$ with $-\omega$ in the integral: while the term $n(-\omega)n(\omega)$ is invariant, the commutator picks up a minus sign.

We would like to extract a number from this operator-valued curvature. Although there is no canonical way to do so\footnote{From a mathematical perspective it corresponds to identifying a suitable dual space of the algebra of zero modes $\mathcal{A}_0$, and corresponding bilinear pairing. In the case of infinite-dimensional algebras this is very subtle (see for example \cite{deBoer:2021zlm} where the case of the Virasoro algebra was discussed).}, a simple and convenient choice is to take the expectation value of the operator $F$ in the original pure state $|\Psi\rangle$:
\beq \label{eq:Fpsi}
F_{\Psi} \equiv\langle \Psi|F| \Psi\rangle=\langle F \rangle~.
\eeq
As we will show in Section~\ref{sec:bulksymp}, it turns out that \eqref{eq:Fpsi} results in the correct identification with the bulk symplectic form. This agreement can be viewed as an argument for why this choice is the most `physical' one. However, from a mathematical point of view we stress that this choice is by no means unique, and the operator $F$ contains more information.

To proceed in evaluating this expectation value, let us mention a well-known result for two-point functions of operators in the global state $|\Psi\rangle$, the so-called KMS condition. (For a pedagogical exposition of the KMS condition, see for example \cite{Witten:2018zxz, Hollands:2022dem}.) Roughly speaking, it says that we can swap operators in a two-point function provided that we evolve one of them in imaginary modular time. To be precise, we introduce the Tomita operator $S_{\Psi}$ as an anti-linear operator that sends 
\beq \label{eq:Tomitaop}
S_{\Psi}\mathcal{O}|\Psi\rangle = \mathcal{O}^{\dagger} |\Psi \rangle~.
\eeq
The modular operator is now defined by $\Delta=S_{\Psi}^{\dagger}S_{\Psi}$, and satisfies $\Delta |\Psi\rangle =|\Psi\rangle$. Using the definition \eqref{eq:Tomitaop} together with anti-linearity one can verify that
\beq 
\langle \Psi | \mathcal{O}\mathcal{O}' | \Psi \rangle = \langle \Psi | \mathcal{O}'\Delta \mathcal{O}|\Psi \rangle~,
\eeq
for $\mathcal{O},\mathcal{O}'\in\mathcal{A}$. One can represent the modular operator in terms of the two-sided modular Hamiltonian $\hat{H}_{\rm mod}\equiv H_{\rm mod}-\bar{H}_{\rm mod}=-\log \Delta$ so that the modular flow \eqref{eq:modularflow} is given by $\mathcal{O}_s=\Delta^{-is}\mathcal{O}\Delta^{is}$. Therefore, assuming that the operators $\mathcal{O}_s(x),\mathcal{O}(x')$ are in the algebra $\mathcal{A}$ associated to the subregion\footnote{To ensure this we need to put a restriction on the support of the sources in the perturbation \eqref{eq:densitymatrixchange}. In the Euclidean picture we assume that the state is perturbed by changing the sources at the branch cut only (using some suitable limiting procedure where we approach it from above and below).} one obtains the condition
\beq \label{eq:KMS}
\langle \mathcal{O}_s(x)\mathcal{O}(x') \rangle = \langle\mathcal{O}(x')\mathcal{O}_{s+i}(x)\rangle~.
\eeq 

The action of modular flow on the Fourier modes $\mathcal{O}_{\omega}$ is particularly simple, i.e., see \eqref{eq:modularevolution}, so that the KMS condition \eqref{eq:KMS} reads
\beq 
\langle\mathcal{O}_{\omega}(x)\mathcal{O}_{\omega'}(x')\rangle =  e^{-\omega}\langle\mathcal{O}_{\omega'}(x')\mathcal{O}_{\omega}(x)\rangle~.
\eeq
By rearranging terms on both sides of the equation one finds the following identity
\beq \label{eq:comm=two-point}
\langle\mathcal{O}_{\omega}(x)\mathcal{O}_{\omega'}(x')\rangle = n(\omega) \langle[\mathcal{O}_{\omega'}(x'),\mathcal{O}_{\omega}(x)]\rangle~,
\eeq
where $n(\omega)$ was defined in \eqref{eq:Boltzmann}. This relation is very useful in practice since we can use it to rewrite the expectation value of a commutator in terms of a two-point function. 

We can now use this to evaluate \eqref{eq:Fpsi}. By recognizing the right-hand side of \eqref{eq:comm=two-point} in $F_\Psi$, we obtain
\beq 
F_{\Psi} =\mathcal{N}^{-1}\int d^dx \int d^dx'\, \delta_1 \lambda(x) \delta_2 \lambda(x') \int d\omega\,n(\omega) \langle \mathcal{O}_{-\omega}(x)\mathcal{O}_{\omega}(x')\rangle~.
\eeq 
One can rewrite the above result by putting one of the two operators in its original form. Using the definition \eqref{eq:frequencyoperators} and the condition $\Delta|\Psi\rangle=|\Psi\rangle$ one can show that the modular Fourier modes satisfy the following relation: 
\begin{align}
\langle \mathcal{O}_{\omega} &(x)\mathcal{O}_{\omega'}(x') \rangle = \int_{-\infty}^{\infty} ds \int_{-\infty}^{\infty} ds' e^{-i(\omega s+\omega's')}\langle \Psi |\mathcal{O}_s(x)\mathcal{O}_{s'}(x') |\Psi\rangle \nonumber \\
&=\int_{-\infty}^{\infty} ds\, e^{-i\omega s}\int_{-\infty}^{\infty} ds' e^{-i(\omega+\omega')s'}\langle \Psi | e^{iH_{\rm mod}s'}\mathcal{O}_s(x)\mathcal{O}(x') e^{-iH_{\rm mod}s'}| \Psi\rangle \nonumber \\
&=\int_{-\infty}^{\infty} ds\, e^{-i\omega s}2\pi \delta(\omega+\omega')\langle \Psi | \mathcal{O}_s(x)\mathcal{O}(x')| \Psi\rangle =2\pi \delta(\omega+\omega')\langle \mathcal{O}_{\omega}(x)\mathcal{O}(x') \rangle~.
\end{align}
Hence, we conclude that the $\delta(0)$ factor drops out of the final answer, and we obtain
\beq  \label{eq:finalmodularBerry}
F_{\Psi} = \frac{1}{2\pi}\int d^dx \int d^dx'\, \delta_1 \lambda(x) \delta_2 \lambda(x') \int d\omega \,n(\omega) \langle \mathcal{O}(x)\mathcal{O}_{\omega}(x')\rangle~.
\eeq 
This last equation will be useful in finding a bulk interpretation for the Berry curvature. But first we will show that one can also extract from the Berry curvature a symmetric quantity, which behaves like an information metric on the space of modular Hamiltonians. 

\subsection{Quantum information metric}\label{sec:QImetric}
We can obtain some additional information from $F$ that will be also useful from the bulk perspective. Specifically, it is convenient to construct a symmetric quantity from $F$ by taking one of the perturbations to be of the form $[H_{\rm mod},X]$. The quantity
\beq \label{eq:Berrymetric}
G=P_0\left([X_1,[H_{\rm mod},X_2]]\right)
\eeq
is symmetric under exchanging $X_1$ with $X_2$. Using an additional commutation $[H_{\rm mod}, \cdot]$ to turn the antisymmetric object $P_0([X_1, X_2])$ into a symmetric one follows a well-known construction, which applies in finite-dimensional settings \cite{andersson}. 

To see how this works, we use the Jacobi identity reorganized in the form
\beq \label{eq:Jacobi}
[X_1,[H_{\rm mod},X_2]] = [X_2,[H_{\rm mod},X_1]]+[H_{\rm mod},[X_1,X_2]]~.
\eeq
Since the last term lies in the image of the adjoint action of $H_{\rm mod}$, it is zero-mode free by \eqref{eq:projprop}, so that
\beq 
P_0([H_{\rm mod},[X_1,X_2]])=0~.
\eeq
Therefore, taking the projection on both sides of \eqref{eq:Jacobi} gives the required relation
\beq 
P_0([X_1,[H_{\rm mod},X_2]])=P_0([X_2,[H_{\rm mod},X_1]])~.
\eeq

Using the fact that $\mathcal{O}_{\omega}$ is an eigenoperator with respect to the adjoint action of $H_{\rm mod}$, ~\eqref{eq:eigenequation}, we pick up an extra factor of $\omega$ when evaluating $[H_{\rm mod},X_2]$. Indeed, the formula for $F$ gets modified to 
\begin{align} \label{eq:G}
G=\mathcal{N}^{-1}\int d^dx \int d^dx'\, \delta_1 \lambda(x) \delta_2 \lambda(x') \int d\omega \, n(-\omega) n(\omega)\omega[\mathcal{O}_{\omega}(x),\mathcal{O}_{-\omega}(x')]~.
\end{align}
Note that this expression is indeed symmetric under the replacement of $\omega$ with $-\omega$. As we did for $F_{\Psi}$, one can extract from $G$ a number by taking an expectation value, $G_{\Psi}\equiv\langle \Psi|G|\Psi\rangle$. Going through a similar set of computations one obtains 
\beq \label{eq:Gpsi}
G_{\Psi}=\frac{1}{2\pi}\int d^dx \int d^dx'\, \delta_1 \lambda(x) \delta_2 \lambda(x') \int d\omega \, \omega\, n(\omega) \langle \mathcal{O}(x)\mathcal{O}_{\omega}(x')\rangle~.
\eeq 

This expression can be rewritten in a form which makes the relation with  quantum information theory manifest. Namely, one can undo the Fourier transformation \eqref{eq:frequencyoperators} and write the integral over modular frequencies in terms of an integral over modular time. The extra factor of $\omega$ in \eqref{eq:Gpsi} comes in handy, since we can replace $\omega n(\omega)$ with the following integral\footnote{This can be derived from an application of the residue formula (by closing the $s$-contour in the upper/lower half plane depending on the sign of $\omega$) and the geometric series relation. In particular, one uses that the residue at $s=ik$ for $k\in \mathbb{Z}$ is given by 
\beq \label{eq:residue}
\mathrm{Res}_{s=ik} \frac{\pi}{\sinh^2(\pi s)} f(s)=\frac{f'(ik)}{\pi}~.
\eeq}:
\beq \label{eq:sinhintegral}
|\omega|\, n(\omega)=\int_{-\infty -i\epsilon}^{\infty -i \epsilon}ds \frac{\pi}{2\sinh^2(\pi s)}e^{-i\omega s}~.
\eeq
Combining \eqref{eq:Gpsi} with \eqref{eq:sinhintegral} and applying the inverse of the Fourier decomposition, \eqref{eq:frequencyoperators}, we find that 
\beq \label{eq:modularBerrymetric}
G_{\Psi} =\int d^dx \int d^dx'\, \delta_1 \lambda(x) \delta_2 \lambda(x') \int_{-\infty -i\epsilon}^{\infty -i \epsilon}ds \frac{\pi}{2\sinh^2(\pi s)} \langle \mathcal{O}(x)\mathcal{O}_s(x')\rangle~.
\eeq
This quantity agrees with a well-known quantum information theoretic `metric' on the space of mixed states \cite{Lashkari:2015hha,Faulkner:2017tkh}, which is obtained from the second variation of the relative entropy. (See Appendix~\ref{sec:relativeentropy} for more details.) We therefore see that the parallel transport problem for modular Hamiltonians is closely related to a metric on the space of density matrices. This should be reminiscent of the similar situation in the case of pure states, where the Berry phase computes the Fubini-Study metric on the space of pure states \cite{Belin:2018bpg,Belin:2018fxe}. It also provides a natural starting point for investigations of a bulk interpretation.

\section{Relation to the bulk symplectic form}\label{sec:bulksymp}

We would now like to derive a bulk interpretation of the modular Berry curvature and information metric. Let us start out by defining a quantity that generalizes both \eqref{eq:finalmodularBerry} and \eqref{eq:Gpsi}:
\beq \label{eq:HH}
H_{\Psi}=\frac{1}{2\pi}\int d^dx \int d^dx'\, \delta_1\lambda(x) \delta_2\lambda(x') \int d\omega \, \mathcal{F}(\omega) \langle \mathcal{O}(x)\mathcal{O}_{\omega}(x')\rangle~.
\eeq
$H_\Psi$ reproduces $F_{\Psi}$ for the choice $\mathcal{F}(\omega) = n(\omega)$, and $G_{\Psi}$ for $\mathcal{F}(\omega) = \omega \,n(\omega)$. 

\begin{figure}[t!]
    \centering
    \includegraphics[width=0.58\columnwidth]{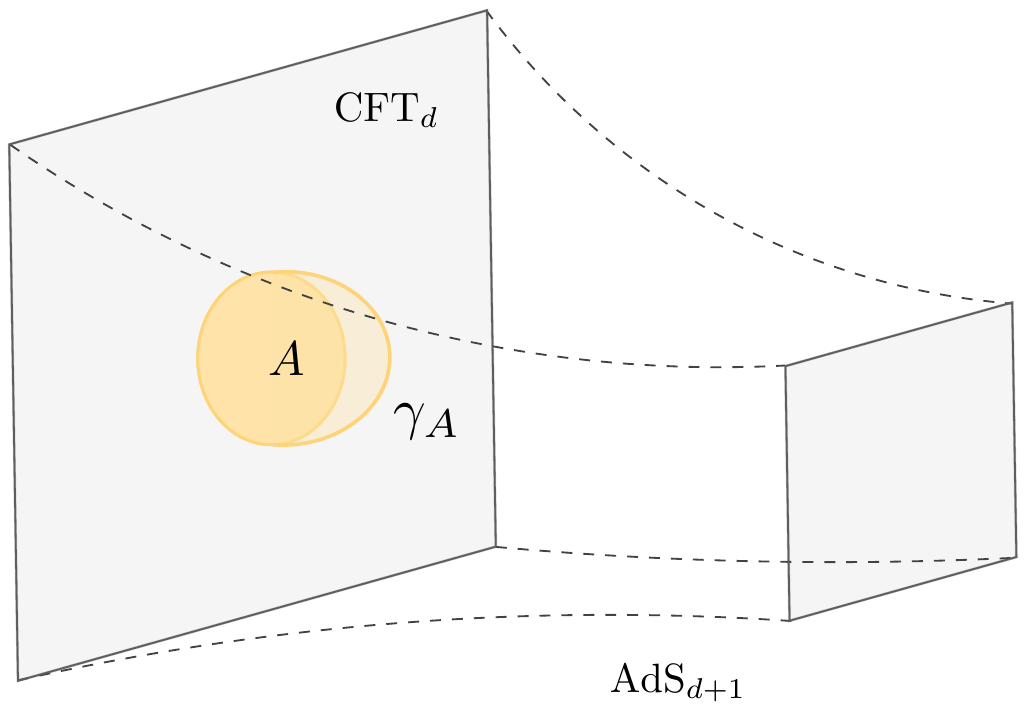}
    \caption{A depiction of the entanglement wedge $\Sigma=\Sigma_A$ (semi-transparent yellow) on a fixed time slice. Its boundary $\partial \Sigma_A$ has two components: the region $A$ at the asymptotic boundary and the Ryu-Takayanagi (RT) surface $\gamma_A$ that extends into the bulk AdS spacetime.}
    \label{fig:entanglementwedge}
\end{figure}

We would like to extend $H_\Psi$ into the bulk. Let us for the moment assume that the boundary operator $\mathcal{O}$ used to deform the state in \eqref{eq:Sdeform} is a scalar of conformal dimension $\Delta_+$. By general AdS/CFT principles, the dual bulk description is some scalar operator $\Phi$ localized in a spacelike slice $\Sigma=\Sigma_A$ of the entanglement wedge associated to the boundary region $A$ (see Figure \ref{fig:entanglementwedge}). Since the expression \eqref{eq:HH} contains operators of the form $\mathcal{O}_{\omega}$, we first need to describe the bulk analogue of the modular Fourier decomposition. Using the equivalence of bulk and boundary modular flows it is then possible to extend the modular frequency modes into the bulk~\cite{Jafferis:2015del, Faulkner:2017vdd, May:2018tir}. We will argue that the two-point function in \eqref{eq:HH} behaves like the asymptotic flux of some suitably defined symplectic form. Its bulk extension provides a natural definition for the bulk symplectic form associated to the entanglement wedge.
A similar approach was used in \cite{May:2018tir} to find holographic duals of the $\alpha$-$z$ relative R\'enyi divergences, which are certain generalizations of relative entropy.

Note that while we will focus explicitly on scalar case in this section, it is straightforward to extend to more general field deformations. An important generalization to stress tensor insertions, which in the bulk correspond to perturbations of the geometry, will be treated in Section~\ref{sec:GravSympForm}.

\subsection{Bulk phase space}
To make the computation tractable we will make one further assumption, namely that we are working in a free field approximation. We expect this approximation to hold for a generic bulk quantum theory to leading order in $1/N$. Moreover, we are interested in the symplectic form evaluated at a particular point in phase space, so in principle it is possible to find bulk variables $\Phi$ and $\Pi$ that linearize the symplectic form.  Interactions can be included perturbatively. 

The bulk phase space can now be described explicitly in terms of the operators $\Phi(X)$ and canonically conjugate operators $\Pi(X)$ for $X\in \Sigma$. They satisfy the canonical commutation relations 
\begin{equation} \label{eq:commutationrelations}
    [\Phi(Y),\Pi(X)]=i\delta(X-Y)~.
\end{equation}
Let us again now introduce the bulk modular flow associated to $\Sigma$, implemented by some bulk density matrix $\rho_{\rm bulk}$. Following a similar procedure as in the boundary CFT one can now decompose the operators $\Phi$ into modular Fourier modes 
\begin{equation} \label{eq:Phiomega}
    \Phi_{\omega}=\int_{-\infty}^{\infty} ds\, e^{-i\omega s} \rho_{\rm bulk}^{-is} \Phi\rho_{\rm bulk}^{is}~,
\end{equation}
and similarly for $\Pi$. Given that the operators $\{\Phi(X), \Pi(X)\, |\, X\in \Sigma \}$ are a formal basis for $\mathcal{A}_{\Sigma}$, we can express the operator \eqref{eq:Phiomega} as the linear combination
\begin{equation} \label{eq:Phiexpansion1}
\Phi_{\omega}(X)=\int_{\Sigma} dY\left[\alpha_{\omega}(X,Y)\Phi(Y)+\beta_{\omega}(X,Y)\Pi(Y) \right]~,
\end{equation}
with $\alpha_{\omega}(X,Y),\beta_{\omega}(X,Y)$ coefficients in the expansion. By acting with the commutator on \eqref{eq:Phiexpansion1} and taking the expectation value in the state $\rho_{\rm bulk}$ we find that 
\begin{equation}
\alpha_{\omega}(X,Y) = i  \langle [\Pi(Y),\Phi_{\omega}(X)]\rangle = \frac{i}{n(\omega)} \langle \Phi_{\omega}(X)\Pi(Y) \rangle~,
\end{equation}
where we have used the KMS condition \eqref{eq:comm=two-point} adapted to the bulk correlation function. Similarly we have,
\begin{equation}
\beta_{\omega}(X,Y) = - i  \langle [\Phi(Y),\Phi_{\omega}(X)]\rangle = -\frac{i}{n(\omega)} \langle \Phi_{\omega}(X)\Phi(Y) \rangle~.
\end{equation}
By plugging this into \eqref{eq:Phiexpansion1} one obtains the final result 
\begin{equation} \label{eq:Phiexpansion2}
\Phi_{\omega}(X)=\frac{i}{n(\omega)}\int_{\Sigma} dY\left[\langle \Phi_{\omega}(X)\Pi(Y) \rangle\Phi(Y)-\langle \Phi_{\omega}(X)\Phi(Y) \rangle\Pi(Y) \right]~.
\end{equation}
One can also view \eqref{eq:Phiexpansion2} as a Bogoliubov transformation, which changes the operator basis from $\Phi,\Pi$ to modular Fourier modes. As mentioned in Section~\ref{sec:BerryCurv}, the advantage of using the operators $\Phi_{\omega}$ is that in this basis the action of the (bulk) modular Hamiltonian is relatively simple. Note that to completely specify the right-hand side of \eqref{eq:Phiexpansion2} requires some boundary condition at the finite boundary of $\Sigma$, i.e., at the RT surface (see Figure \ref{fig:entanglementwedge}). We will come back to this issue later, and will argue that the behavior of the integrand near the RT surface is related to the presence of a zero mode in the Berry transport problem.

\subsection{Modular extrapolate dictionary}
Up to this point, we have only used some basic properties of the bulk operator algebra in a free field approximation to write \eqref{eq:Phiexpansion2}. Let us now invoke the AdS/CFT dictionary to relate the operator $\Phi_{\omega}$ to the corresponding boundary operator $\mathcal{O}_{\omega}$. We denote the holographic direction of the AdS space by $z$, so that the bulk coordinate is given by $X=(z,x)$. The extrapolate dictionary now states that the properly regularized version of $\Phi$ approaches the operator $\mathcal{O}$ near the asymptotic boundary:
\begin{equation}
    \lim_{z\to 0} z^{-\Delta_+}\Phi(x,z)=\mathcal{O}(x)~.
\end{equation}
Since we are interested in the modular frequency modes, we will need to use a version of the extrapolate dictionary that is suited to this decomposition. A crucial result was given in \cite{Jafferis:2015del}, where the authors show that the bulk and boundary modular flows agree to first order in $1/N$. This can be used to derive the so-called modular extrapolate dictionary \cite{Faulkner:2017vdd}:
\begin{equation} \label{eq:modularextra}
    \lim_{z\to 0} z^{-\Delta_+}\Phi_{\omega}(x,z)=\mathcal{O}_{\omega}(x)~.
\end{equation}
We will use \eqref{eq:modularextra} to extend the operator $H_{\Psi}$ into the bulk. Indeed, we can take the boundary limit on both sides of the equation in \eqref{eq:Phiexpansion2}:
\begin{equation} \label{eq:bulkOomega}
\mathcal{O}_{\omega}(x)=\frac{i}{n(\omega)}\int_{\Sigma} dY\left[\langle \mathcal{O}_{\omega}(x)\Pi(Y) \rangle\Phi(Y)-\langle \mathcal{O}_{\omega}(x)\Phi(Y) \rangle\Pi(Y) \right]~.
\end{equation}
This formula provides a bulk expression for the boundary operator $\mathcal{O}_{\omega}(x)$ in terms of some bulk-to-boundary propagators. Note that \eqref{eq:bulkOomega} is non-local expression in the bulk, which is a reflection of the non-locality of the action of the modular flow. We can now plug \eqref{eq:bulkOomega} into \eqref{eq:HH} to obtain 
\begin{align} \label{eq:HPsi}
H_{\Psi}=&\,\frac{1}{2\pi}\int_{\Sigma} dY \int d^dx \int d^dx'\, \delta_1\lambda(x) \delta_2\lambda(x') \int d\omega \, \mathcal{C}(\omega) \nonumber \\
& \hspace{20pt}\times \left[\langle \mathcal{O}_{\omega}(x')\Pi(Y) \rangle \langle \mathcal{O}(x)\Phi(Y)\rangle-\langle \mathcal{O}_{\omega}(x')\Phi(Y) \rangle\langle \mathcal{O}(x)\Pi(Y)\rangle \right]~.
\end{align}
We have collected the additional dependence on the modular frequency $\omega$ in the function $\mathcal{C}(\omega)$. It is given by
\begin{equation} \label{eq:relCF}
    \mathcal{C}(\omega)\equiv i\mathcal{F}(\omega)n(\omega)^{-1}~.
\end{equation}
The expression \eqref{eq:HPsi} takes a very simple form when written in terms of the bulk fields. Note that the bulk density matrix $\rho_{\rm bulk}$ gets perturbed in a similar way as the boundary density matrix \eqref{eq:densitymatrixchange}. From the coherent state deformation
\begin{equation}
    \delta \rho_{\rm bulk} = - \int d^{d} x \, \rho_{\rm bulk} \delta\lambda(x) \mathcal{O}(x)~,
\end{equation}
we find that the expectation value of the operator $\Phi$ in the perturbed density matrix $\delta \rho_{\rm bulk}$ is given by
\begin{equation} \label{eq:deltaphii}
\delta \phi (Y) \equiv -\int d^d x \,\delta\lambda(x)\langle\mathcal{O}(x)\Phi(Y) \rangle~. 
\end{equation}
Note that $\delta \phi$ is a number, while $\Phi$ is an operator. Using again that the bulk and boundary modular flows agree we also obtain the relation 
\begin{equation} \label{eq:deltaphiiomega}
\delta \phi_{\omega}(Y)=\int d^d x \,\delta\lambda(x)\langle\mathcal{O}(x)\Phi_{\omega}(Y) \rangle=\int d^d x \,\delta\lambda(x)\langle\mathcal{O}_{-\omega}(x)\Phi(Y) \rangle~. 
\end{equation}
Introducing similar expressions for the canonical conjugate bulk fields $\delta \pi$ and $\delta \pi_{\omega}$ defined in terms of $\Pi$ and $\Pi_{\omega}$ one finds that \eqref{eq:HPsi} simplifies to  
\begin{align} \label{eq:HH2}
H_{\Psi}= \frac{1}{2\pi}\int_{\Sigma} dY \int d\omega \, \mathcal{C}(\omega) \left[\delta_2\pi_{-\omega}(Y)\delta_1\phi(Y)-\delta_2\phi_{-\omega}(Y) \delta_1\pi(Y) \right]~.
\end{align}
The variations $\delta_{1,2}$ that we introduced in the above expression correspond to the choice of sources $\delta_{1,2}\lambda$ respectively in \eqref{eq:densitymatrixchange}. 

\subsection{Entanglement wedge symplectic form}
As a final step we will now perform the integral over modular frequencies to obtain the bulk symplectic form. A convenient trick is to first replace the $\omega$-integral by the action of a suitable differential operator on the bulk fields~\cite{May:2018tir}. Specifically, starting with the modular Fourier modes for an arbitrary function $f$
\beq
f_{-\omega} = \int_{-\infty}^{\infty}ds \, e^{i\omega s}f_s~,
\eeq
we can apply an integration by parts on a wavepacket of such modular Fourier modes to obtain
\begin{align}
    \int d\omega \, \mathcal{C}(\omega) f_{-\omega}(Y)
    =\int_{-\infty}^{\infty} ds\, \int d\omega \, f_s(Y)\mathcal{C}(-i\partial_s) e^{i\omega s}
     =2\pi\int_{-\infty}^{\infty} ds\, \left( \mathcal{C}(i \partial_s)f_s(Y) \right) \delta(s)~.
\end{align}
We thus have 
\beq \label{eq:Chat}
\int d\omega \, \mathcal{C}(\omega) f_{-\omega}(Y) = 2\pi \left( \hat{\mathcal{C}}\,f_s(Y) \right) \Big|_{s=0}~,
\eeq
where we have defined a differential operator $\hat C$ that acts on the modular time $s$ as $\hat{\mathcal{C}}=\mathcal{C}(i\partial_s)$. 

Now we have all the ingredients necessary to match our expression for $H_\Psi$ to a bulk symplectic form. Formally, we define the entanglement wedge symplectic form in terms of the field perturbations and corresponding canonical momenta by 
\begin{equation}
\Omega(\delta_1\phi,\delta_2\phi)=\int_{\Sigma} dY \left[\delta_1\phi(Y)\delta_2\pi(Y)- \delta_1\pi(Y)\delta_2\phi(Y) \right]~.
\end{equation}
Note that $\Omega(\delta_1\phi,\delta_2\phi)$ is manifestly anti-symmetric under interchanging $1$ with $2$. Combining \eqref{eq:HH2} with \eqref{eq:Chat} we find that 
\begin{equation} \label{eq:Homega}
    H_{\Psi}= \Omega(\delta_1\phi,\hat{\mathcal{C}}(\delta_2\phi)_s|_{s=0})~.
\end{equation}

The modular Berry curvature $F_{\Psi}$ in \eqref{eq:finalmodularBerry} is a particular case of this general relation. Recall that the curvature is described by $F_\Psi$, which is obtained from $H_{\Psi}$ by taking the constant function $\mathcal{C}(\omega)=i$. From \eqref{eq:relCF}, this corresponds to the choice $\mathcal{F}=n(\omega)$ in $H_\Psi$. Then from \eqref{eq:Homega}, the modular Berry curvature is exactly proportional to the bulk symplectic form:
\begin{equation} \label{eq:Fbulk}
F_{\Psi} = i \, \Omega(\delta_1\phi,\delta_2\phi)~.
\end{equation}
Note that the factor of $i$ comes from the canonical commutation relations, \eqref{eq:commutationrelations}. The above equality constitutes the main result of this section. It provides a bulk dual for the boundary modular Berry curvature.

The modular Berry metric $G_{\Psi}$ derived in \eqref{eq:Gpsi} also arises as an example of $H_{\Psi}$ with the function $\mathcal{C}(\omega)=i\omega$. In this case, the differential operator acts non-trivially, as $\hat{\mathcal{C}} = - \partial_s$, which corresponds to the infinitesimal action of bulk modular flow. We therefore find that the Berry metric is equal to the bulk symplectic form with an extra action of the modular flow on one of the variations: 
\begin{equation} \label{Gbulk}
G_{\Psi} = \Omega(\delta_1\phi,-\partial_s(\delta_2\phi)_s|_{s=0})~.
\end{equation}
The presence of bulk modular flow in \eqref{Gbulk} can be linked to the extra insertion of the action of the modular Hamiltonian in defining \eqref{eq:Berrymetric}. One can also reverse the logic and argue that the Berry curvature $F_{\Psi}$ via \eqref{eq:Fbulk} provides a natural symplectic form on the space of modular Hamiltonians that agrees with the bulk symplectic form.

Let us now come back to the contribution from the RT surface $\gamma_A$ in \eqref{eq:bulkOomega}. The contribution that is localized on the RT surface is related to the zero mode of the operator $\mathcal{O}$. From the bulk perspective this is quite easy to see: the action of the bulk modular flow leaves the RT surface fixed so, in particular, operators localized at the RT surface commute with the modular Hamiltonian, i.e., they correspond to modular zero modes. 
(See Section~4 of \cite{Faulkner:2017vdd} for an explicit expression of the zero mode $\mathcal{O}_{0}$ in terms of an integral over the RT surface $\gamma_A$ in the case that $\mathcal{O}$ is a scalar.) In our derivation relating the Berry curvature to the bulk symplectic form, we therefore see that the boundary term corresponds to the $\omega=0$ part of the integral over modular frequencies in \eqref{eq:finalmodularBerry}. But this term comes from the zero mode in the original transport operator $X$ as computed in \eqref{eq:Xoperator}. Therefore, imposing $P_0(X)=0$ by subtracting the zero mode from it, and fixing the zero mode ambiguity in the boundary parallel transport problem, naturally fixes the ambiguity in the boundary condition for the entanglement wedge symplectic form to be Dirichlet.

\section{Explicit examples}\label{sec:explicitex}
We will now give some explicit examples that illustrate the formalism we have introduced, but restricted to the scenarios where our state transformation are suitable for describing \emph{shape} transformations. In Section~\ref{sec:stresstensordef}, we will consider the case where the  perturbing operator $\mathcal{O}$ in the Euclidean path integral (see Section~\ref{sec:EuclPI}) is given by a stress tensor deformation. Such a deformation will in general cause a change of the boundary metric, so that it lies in the class of \emph{state}-changing transformations. However, for a particular choice of deformation, namely one generated by a conformal Killing vector, this instead implements a change of \emph{shape} of the entangling surface. From the bulk perspective, this example also illustrates how the derivation of the bulk symplectic form given in Section~\ref{sec:bulksymp} straightforwardly generalizes beyond scalar operators (see Appendix~\ref{sec:gravsympshape} for an explicit evaluation of this symplectic form in the special case of shape deformations).

In Section~\ref{sec:coadjoint}, we will describe shape deformations in terms of symmetries rather than using the Euclidean path integral, which connects to the language of~\cite{deBoer:2021zlm}. We will explain how in the particular case of shape-changing deformations, the Berry curvature is equal to the symplectic form on a special geometry known as a coadjoint orbit. This is reminiscent of the group theoretic structure that was uncovered in two dimensions~\cite{Penna:2018xqq, deBoer:2021zlm}. However, we emphasize that the connection to coadjoint orbits will not carry over in the more general state-changing case.

\subsection{Stress tensor insertions} \label{sec:stresstensordef}

We first consider a specific version of \eqref{eq:densitymatrixchange} where we perturb the state by an insertion of the CFT stress tensor. This class of transformations includes the special case of a state transformation that implements a change of shape of the subregion~\cite{Faulkner:2015csl, Faulkner:2016mzt, Lewkowycz:2018sgn} but also includes non-trivial changes to the boundary metric. In the bulk dual, this will involve perturbations of the gravitational field. 

For concreteness, we consider a $d$-dimensional CFT on the plane $\mathbb{R}^{1,d-1}$ in some global state $|\Psi \rangle$ with some ball-shaped region $A$ in the $t=0$ slice. We consider the modular Hamiltonian associated to some reduced state $\rho=\rho_A$ that is obtained by tracing out the complement of the ball-shaped region $A$. In general, the modular Hamiltonian is a complicated non-local operator, but in the case that $|\Psi\rangle=|0\rangle$ is the vacuum state it has an explicit local expression. One can write
\beq 
H_{\rm mod} = \int_{A} dS^{\nu}\xi_{A}^{\mu}T_{\mu\nu}~,
\eeq
where $T_{\mu\nu}$ is the CFT stress tensor and $\xi_A$ is the vector field that generates modular flow; in particular, it preserves the causal diamond $\mathcal{D}(A)$ of the region $A$. 
We would like to deform the modular Hamiltonian via the the action $\xi$ of some coordinate transformation 
\beq 
x^{\mu} \mapsto x'^{\mu}= x^{\mu}+\xi^{\mu}(x)~.
\eeq 
One can show that the action of $\xi$ is implemented by the action of some operator on the modular Hamiltonian
\beq \label{eq:deltaHmod}
\delta_{\xi} H_{\rm mod}-P_0(\delta_{\xi} H_{\rm mod})=[X,H_{\rm mod}]~,
\eeq
where $X$ is defined by \beq \label{eq:X}
X = \int_{A}dS^{\nu} \xi^{\mu} T_{\mu\nu}~. 
\eeq
The transport problem \eqref{eq:deltaHmod} is in fact a special example of the coherent state formalism that we discussed in Section~\ref{sec:coherentstate}, where we now take $\mathcal{O}=T^{\mu\nu}$ and $\lambda=\partial_{\mu}\xi_{\nu}$.
Before going into the details, we stress that the equality in \eqref{eq:deltaHmod} is actually quite subtle. A general coordinate transformation $\xi$ does not leave the metric $h_{\mu\nu}$ of the CFT invariant. Instead, we have 
\beq 
\delta h_{\mu \nu} = \mathcal{L}_{\xi}h_{\mu\nu} \equiv \partial_{\mu}\xi_{\nu}+\partial_{\nu}\xi_{\mu}~.
\eeq
If $\xi$ is a Killing vector we have $\delta h_{\mu \nu}=0$, but in general the variation is non-zero. The idea is, that for a generic transformation $\xi$, the change in metric can be traded for a change in the state of the CFT implemented by some unitary on the Hilbert space. For this reason, we are able to utilize the formalism of state-changing transformations developed earlier in the paper to describe \emph{shape} changes by restricting to the particular case where $\xi^\mu$ is a conformal Killing vector.

\subsubsection{Deforming the boundary metric} \label{sec:diffeo}
We would like to derive the parallel transport equation, \eqref{eq:deltaHmod}, for this special case of stress tensor insertions. The following subsection will review some results derived in~\cite{Faulkner:2015csl}, while adapting them to the modular Berry setup.  

Under a change of the metric the action of the theory picks up a piece of the form
\beq \label{eq:deltaSboundary}
\delta S \sim \int d^dx \,\delta h_{\mu\nu}(x) T^{\mu\nu}(x)~.
\eeq
Hence, we can think of the deformed state as being obtained from the original state by introducing a source for the stress tensor. We take \eqref{eq:deltaHmod2} as a starting point with the appropriate source and operator. Using a version of the integral formula \eqref{eq:sinhintegral}, one can write this as 
\begin{align}
\langle \omega | & \delta H_{\rm mod} |\omega' \rangle=\int_{-\infty -i\epsilon}^{\infty -i \epsilon}ds\frac{\pi}{2\sinh^2(\pi s)}\int d^dx\, \delta h_{\mu\nu}(x) \langle \omega|\rho^{i s}T^{\mu\nu}(x) \rho^{-i s}|\omega'\rangle~.
\end{align}
The above formula is true for arbitrary metric deformations. Let us now specialize to the case where it is generated by a diffeomorphism: 
\beq
\delta h_{\mu\nu} = \partial_{\mu}\xi_{\nu}+\partial_{\nu}\xi_{\mu}~.
\eeq
We split the integral over the Euclidean plane into two pieces: a tubular neighborhood $R_{b}$ of width $b$ around the entangling region $\partial A$, and its complement $\tilde{R}$. Let us first do the integral over $\tilde{R}$. It can be localized to an integral over the boundary $\partial \tilde{R}$ using an integration by parts:
\begin{align} \label{eq:integrationbyparts}
\int_{\tilde{R}} d^dx\, \partial_{\mu}\xi_{\nu}T^{\mu\nu} = - \int_{\tilde{R}} d^dx\, \xi_{\nu} & \partial_{\mu}T^{\mu\nu}+\int_{\partial \tilde{R}} dS_{\mu}\, \xi_{\nu} T^{\mu\nu}~.
\end{align} 
By conservation of the stress energy on the support of the diffeomorphism, only the second term in \eqref{eq:integrationbyparts} survives. Let us therefore consider the boundary of $\tilde{R}$ that consist of three parts $\partial \tilde{R}=\partial R_{b}\cup \partial \tilde{R}_+ \cup \partial \tilde{R}_{-}$, i.e., the boundary of the tubular neighborhood, and the region just above and below the branch cut at $A$. We first consider the term coming from the branch cut: 
\begin{align}\label{eq:dHmodT}
\delta H_{\rm mod}&\big|_{\rm cut} = \int_{-\infty -i\epsilon}^{\infty -i \epsilon} ds\frac{\pi}{2\sinh^2(\pi s)}\left( \int_{\partial \tilde{R}_+} -\int_{\partial \tilde{R}_{-}} \right)dS_{\mu}\, \xi_{\nu} T^{\mu\nu}_{-s}~.
\end{align}
Here the modular-evolved stress tensor is defined according to \eqref{eq:modularflow} by $T^{\mu\nu}_{-s} = \rho^{is}T^{\mu\nu}\rho^{-is}$. To perform the integral over the branch cut we note that the value of the stress tensor above and below the branch cut are related by modular evolution in Euclidean time. (Recall that Euclidean modular evolution acts geometrically by circular flow around the branch points.) Therefore, we can change the integration region from $\partial \tilde{R}_-$ to  $\partial \tilde{R}_+$ by applying a substitution $s\to s +i -2 i \epsilon$ 
in the integral over $s$. Hence, it follows that:
\begin{align}
\delta H_{\rm mod}&\big|_{\rm cut} =  \int_{-\infty}^{\infty}ds\left(\frac{\pi}{2\sinh^2(\pi(s+i \epsilon))}-\frac{\pi}{2\sinh^2(\pi(s-i \epsilon))} \right)\int_{\partial \tilde{R}_+}dS_{\mu}\, \xi_{\nu} T^{\mu\nu}_s~.
\end{align}
Since the contour now only encloses the pole at zero, the integral over $s$ now precisely picks up the double pole at $s=0$. 
From \eqref{eq:residue} we find that the residue at $s=0$ is given by 
\beq 
\frac{1}{2\pi}\frac{d}{ds}\Big|_{s=0} T^{\mu \nu}_s = \frac{i}{2\pi}[H_{\rm mod},T^{\mu\nu}]~.
\eeq
In the limit $b\to 0$, the region $\partial \tilde{R}_+$ becomes equal to the subregion $A$, so we conclude that:
\beq 
\delta H_{\rm mod}\big|_{\rm cut} =-\int_{A}dS_{\mu}\, \xi_{\nu}[H_{\rm mod}, T^{\mu\nu}]~.
\eeq
This already reproduces the result \eqref{eq:deltaHmod}. For a detailed derivation of the corner term contribution from $\partial R_b$, see~\cite{Faulkner:2015csl}. For our purposes, we will neglect this term since it is unaffected by a shift in modular time (which is how the modular Hamiltonian acts close to the boundary of the subregion). Thus, it will commute with the modular Hamiltonian and therefore only contributes to the zero mode piece $P_0(\delta H_{\rm mod})$ in the modular transport problem, and will not affect the Berry curvature. 

\subsubsection{Gravitational bulk symplectic form}\label{sec:GravSympForm}
Since the stress tensor perturbations on the boundary are related to perturbations of the bulk geometry, we will compute the gravitational bulk symplectic form explicitly, and compare to the result obtained from the Berry curvature. A standard way to compute the bulk symplectic form is using the covariant phase space formalism \cite{Crnkovic:1986ex,Iyer:1995kg,Iyer:1994ys,Harlow:2019yfa}. This starts from the general action
\be
S=\int L~,
\ee
where $L$ is the Lagrangian density which is a $(d+1)$-form on spacetime. We follow standard conventions and denote the exterior derivative on field space by $\delta$, and the exterior derivative on spacetime by $d$.  

\begin{figure}
    \centering
    \includegraphics[width=0.6\columnwidth]{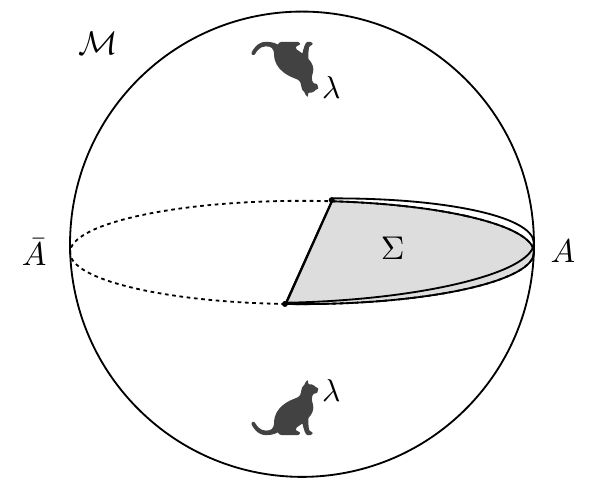}
    \caption{The location of the entanglement wedge $\Sigma$ (gray) in the ball-shaped Euclidean bulk geometry that is used to prepare the state $\rho_{\rm bulk}$. The boundary of $\Sigma$ consists of the subregion $A$ on the boundary of the ball, and the RT surface that extends through the bulk.}
    \label{fig:entanglementbulk}
\end{figure}

We write the variation of the Lagrangian as 
\be
\delta L = E\, \delta \varphi + d\Theta~,
\ee
where $\varphi$ denotes the collection of dynamical fields of the theory, and $E$ are the equations of motion (which vanish on-shell). The boundary term $\Theta$ is a one-form on field space and a $d$-form on spacetime.
Its variation $\omega \equiv \delta \Theta$ is a two-form on field space that can be integrated to give a symplectic form:
\be \label{eq:covariantphasespace}
\Omega =  \int_{\Sigma} \omega~.
\ee
The $d$-dimensional surface $\Sigma$ is usually taken to be a complete Cauchy surface of the bulk spacetime. In our case, we will be interested in the situation where $\Sigma$ only covers part of the Cauchy slice that corresponds to the entanglement wedge of some boundary subregion (see Figure \ref{fig:entanglementbulk}).

Let us now consider the case of pure Einstein gravity with Lagrangian 
\be
L=\frac{1}{16\pi G}\left(R-2\Lambda\right)\epsilon~,
\ee
where $\epsilon$ is the $(d+1)$-dimensional volume form. We will take $g_{\alpha \beta}$ to be the bulk metric. It is straightforward to show that 
$\Theta$ takes the form 
$\Theta=\theta \cdot \epsilon$
with 
\be \label{eq:thetaalpha}
\theta_{\alpha}=\frac{1}{16\pi G}g^{\beta \gamma} \left(\nabla_{\gamma}\delta g_{\alpha \beta}-\nabla_{\alpha}\delta g_{\beta \gamma} \right) ~.
\ee
Let us now compute the pullback of $\theta$ to the surface $\Sigma$. Denoting by $n_a$ the unit normal vector to $\Sigma$, and $\epsilon_{\Sigma}$ the associated volume form, one finds that 
\begin{align} \label{eq:normaltimestheta}
    n^{\alpha} \theta_{\alpha} = \frac{1}{16\pi G}\left(n_\rho g^{\beta \sigma} - n^{\sigma} \delta^{\beta}_\rho \right) \delta \Gamma^{\rho}_{\sigma \beta}
      = \frac{1}{16\pi G}\left(n_\rho \gamma^{\beta \sigma} - n^{\sigma} \delta^{\beta}_\rho \right) \delta \Gamma^{\rho}_{\sigma \beta}~,
\end{align}
where we have introduced the induced metric  $\gamma_{\alpha\beta}= g_{\alpha\beta} - n_{\alpha} n_{\beta}$ on $\Sigma$. 
Then the pullback of $\Theta$ to $\Sigma$ can be written in terms of the extrinsic curvature as 
\cite{Harlow:2019yfa, Burnett:1990tww}
\be \label{pullbackfullexpression}
\Theta|_{\Sigma}=n_{\alpha}\theta^{\alpha} \epsilon_{\Sigma} = \delta \left(-\frac{1}{8\pi G} K \epsilon_{\Sigma}  \right) - \frac{1}{16\pi G} \left(K^{\alpha\beta} - K \gamma^{\alpha\beta} \right) \delta \gamma_{\alpha \beta} \epsilon_{\Sigma}  +  dC~.
\ee
The term $C = c \cdot \epsilon_{\Sigma}$ is given by 
\be
c^{\alpha} = -\frac{1}{16\pi G} \gamma^{\alpha \beta} n^{\rho} \delta g_{\beta \rho}~,
\ee
and vanishes if we impose Dirichlet boundary conditions. In that case, we can express  \eqref{pullbackfullexpression} in terms of the quantities
\be 
\pi^{\alpha\beta} \equiv -\frac{1}{16 \pi G} \left(K^{\alpha\beta} - K \gamma^{\alpha\beta} \right)~, \hspace{20pt} \ell \equiv \frac{1}{8\pi G} K \epsilon_{\Sigma}~.
\ee
Indeed, we have
\be \label{eq:pullbackTheta}
\Theta|_{ \Sigma}=\pi^{\alpha\beta}\delta \gamma_{\alpha\beta}\epsilon_{ \Sigma} -\delta \ell~.
\ee
The fields $\pi^{\alpha\beta}$ will play the role of the canonical momenta associated to the induced metric.  Finally, taking another variation of \eqref{eq:pullbackTheta} one finds that 
\beq
\delta \Theta|_{\Sigma}= \left(\delta\pi^{\alpha\beta}\wedge \delta \gamma_{\alpha\beta}\right)\epsilon_{\Sigma}~.
\eeq
This leads to the final expression for the bulk symplectic form in Darboux form: 
\be \label{eq:BULKSYMP}
\Omega(\delta_1g,\delta_2g)=
\int_{\Sigma}dX \left[\delta_1 \pi^{\alpha\beta
}\delta_2 \gamma_{\alpha\beta} - \delta_2 \pi^{\alpha\beta}\delta_1 \gamma_{\alpha\beta}\right]~.
\ee

The boundary quantity \eqref{eq:HH} that comes from the Berry transport problem in the case of stress tensor deformations is given by
\begin{align}\label{eq:gravsusceptibilty}
H_{\Psi}=\frac{1}{2\pi}\int d\omega \, \mathcal{F}(\omega) \int d^d x \int d^d x'\, \delta_1 \lambda_{\mu\nu}(x) \delta_2 \lambda_{\sigma \tau}( x') \langle T^{\mu\nu}(x)T^{\sigma\tau }_{\omega}(x') \rangle~,
\end{align}
where $\delta \lambda_{\mu\nu}(x)$ is generated by a change of boundary metric as in \eqref{eq:deltaSboundary}. Let us now compare \eqref{eq:BULKSYMP} to the Berry curvature. The computation is very similar to that of the scalar field, with the difference that some extra indices appear. We denote the bulk operator corresponding to the induced metric $\gamma_{\alpha\beta}$ by $\Gamma_{\alpha\beta}$, and its canonical conjugate operator by $\Pi_{\alpha\beta}$. The commutation relations are analogous to the scalar field case \cite{May:2018tir}: 
\begin{align}
[\Gamma_{\alpha\beta}(X),\Pi_{\sigma \tau}(Y)] = \frac{i}{2}(\delta_{\alpha\sigma}\delta_{\beta\tau}+\delta_{\alpha\tau}\delta_{\beta\sigma})\delta(X-Y)~.
\end{align}
As before, we can define the modular Fourier modes $\Gamma^{\omega}_{\alpha\beta}$ associated to the operator $\Gamma_{\alpha\beta}$, and expand in terms of $\Gamma_{\alpha\beta}, \Pi_{\alpha \beta}$.
Similarly to before, the coefficients can be written in terms of two-point functions using the KMS condition. Applying a version of the modular extrapolate dictionary \eqref{eq:modularextra}  that is suited to metric perturbations 
one finds that 
\begin{align} \label{eq:somerelation}
T_\omega^{\mu\nu} (x) = \frac{i}{n(\omega)} \int_{\Sigma} dX \left[ \langle T_\omega^{\mu\nu}(x) \Gamma_{\alpha\beta}(X) \rangle \Pi^{\alpha\beta}(X) - \langle T_\omega^{\mu\nu}(x) \Pi_{\alpha\beta}(X)\rangle \Gamma^{\alpha\beta}(X)\right].
\end{align}
Plugging \eqref{eq:somerelation} into \eqref{eq:gravsusceptibilty}
we find that 
\begin{align}
H_{\Psi}&= \frac{1}{2\pi}\int_{\Sigma}dX \, \int d\omega \, \mathcal{C}(\omega) \int d^d x \int d^d x' \, \delta_1 \lambda_{\mu \nu}(x) 
\delta_2 \lambda_{\sigma \tau}(x') \nonumber \\
&\times \left[ \langle T^{\mu \nu}(x) \Pi^{\alpha \beta}(X)\rangle \langle T_{\omega}^{\sigma \tau}(x') \Gamma_{\alpha\beta}(X)\rangle - \langle T^{\mu\nu}(x) \Gamma^{\alpha\beta}(X)\rangle\langle T_{\omega}^{\sigma \tau}(x') \Pi_{\alpha\beta}(X)\rangle  \right].
\end{align}
Similarly to the scalar field case, one can now write the above expression in terms of the metric perturbations and canonical momenta by evaluating the relevant operator
in the perturbed state. For example, we have an identity of the form 
\beq 
\delta \gamma^{\alpha\beta}_{-\omega} = -  \int d^d x\,\delta \lambda_{\mu \nu}(x)\,\langle T^{\mu\nu}_{\omega}(x) \Gamma^{\alpha\beta}\rangle~.
\eeq
Using this together with similar expressions for the perturbations $\delta \gamma_{\alpha\beta},\delta\pi_{\alpha\beta},\delta \pi^{\alpha \beta}_{-\omega}$, one can write
\begin{align}
 H_{\Psi} = \frac{1}{2\pi}\int_{\Sigma} dX \int d\omega \,\mathcal{C}(\omega) \left[\delta_1\pi_{\alpha\beta}(X) \delta_2 \gamma^{\alpha\beta}_{-\omega}(X) - \delta_1 \gamma_{\alpha\beta}(X) \delta_2\pi_{-\omega}^{\alpha\beta}(X)\right]~.
\end{align}
Applying \eqref{eq:Chat} to remove the integral over frequencies one finds that 
\beq 
H_{\Psi}=\int_{\Sigma}dX \left[\delta_1 \pi_{\alpha\beta
}\,\hat{\mathcal{C}}\left(\delta_2 \gamma^{\alpha\beta}\right)_{s}\Big|_{s=0} - \delta_1 \gamma_{\alpha\beta}\hat{\mathcal{C}}\,\left(\delta_2 \pi^{\alpha\beta}\right)_s\Big|_{s=0}~\right],
\eeq
where the insertion of the operator $\hat{\mathcal{C}}$ is defined in \eqref{eq:Chat}. In the case of the Berry curvature $F_{\Psi}$ where $\mathcal{C}(\omega)=i$ is the constant function, this explicitly agrees with the gravitational bulk symplectic form~\eqref{eq:BULKSYMP}. 

For the symmetric quantity \eqref{eq:Berrymetric}, which results from taking $\mathcal{C}(\omega)=i\omega$ to no longer be constant, one finds that
\beq \label{eq:G=canonicalenergy}
G_{\Psi}=\Omega(\delta_1 g, \mathcal{L}_{\xi}\delta_2g)~,
\eeq
where the bulk modular flow associated to the vacuum state acts via the Lie derivative $\mathcal{L}_{\xi}$. 
This quantity is also known as the canonical energy \cite{Lashkari:2015hha, May:2018tir,Hollands:2012sf}.
From the boundary definition, it is obvious that~\eqref{eq:Berrymetric}  defines a symmetric quantity. To see from the bulk perspective that~\eqref{eq:G=canonicalenergy} is symmetric under the interchange of $1$ and $2$, one can use the product rule and Cartan's magic formula to write
\beq \label{eq:Cartan}
\Omega(\delta_1 g, \mathcal{L}_{\xi}\delta_2g)-\Omega(\mathcal{L}_{\xi}\delta_1 g,\delta_2g) = \int_{\Sigma} \mathcal{L}_{\xi}\delta \Theta = \int_{\Sigma} d(i_{\xi}\delta \Theta)~.
\eeq
At the last equality we have also used that the symplectic potential $\omega=\delta \Theta$ is closed, i.e., $d\omega =0$. Now we can use Stokes' theorem to localize the integral in \eqref{eq:Cartan} to the boundary $\partial \Sigma$, which consists of the RT surface and the asymptotic boundary. Using the fact that the diffeomorphism $\xi$ is an asymptotic Killing vector which vanishes at the RT surface as well, we find that the boundary terms vanish:
\beq 
\Omega(\delta_1 g, \mathcal{L}_{\xi}\delta_2g)-\Omega(\mathcal{L}_{\xi}\delta_1 g,\delta_2g)= \int_{\partial \Sigma} i_{\xi}\delta \Theta =0~.
\eeq
This confirms that the canonical energy is symmetric, following our derivation of \eqref{eq:G=canonicalenergy}.

\subsection{Symmetry transformations} \label{sec:coadjoint}

We will now study the case where the diffeomorphisms that implement the deformation explicitly lie in the conformal group. This is the direct higher-dimensional generalization of the shape-changing setup that was considered in~\cite{Czech:2017zfq, Czech:2019vih} and reviewed in~\cite{deBoer:2021zlm} for the case of AdS$_3$/CFT$_2$. In particular, we will show that the resulting geometric space has the structure of a coadjoint orbit of the conformal group. Notably, the specific state-changing transformations that were considered in \cite{deBoer:2021zlm} are \emph{not} part of the symmetry algebra of $\mathrm{CFT}_d$ when $d>2$, which is finite-dimensional. This is to be contrasted with the situation in $d=2$, where the symmetry algebra is the infinite-dimensional Virasoro algebra. 

\subsubsection{Berry curvature}

Let us consider a CFT$_d$ in the vacuum state. The modular Hamiltonian associated to a spherical region $A$ of radius $R$ is an element the conformal algebra, $\mathfrak{so}(2,d)$. For example, using planar coordinates $(t,\vec{x})$ for the boundary CFT, and choosing a sphere of midpoint $\vec{x}_0$ and radius $R$ in the $t=0$ slice, $H_{\rm mod}$ is generated by the conformal Killing vector that preserves a diamond, which is given by~\cite{deBoer:2016pqk}
\be H_{\rm mod} = \frac{\pi}{R}\left[(R^2-|x-x_0|^2 -t^2)\partial_t -2t(x^i - x_0^i)\partial_i\right]~.
\ee
Using the conventions of Appendix~\ref{appendixA}, we can write this operator in terms of the conformal group generators as
\be \label{eq:Hmod}
H_{\rm mod} = \frac{\pi}{R} \left[-(R^2 - |\vec{x}_0|^2 )P_0 - 2 x_0^i M_{0i}-C_0 \right]~.
\ee

A crucial ingredient in the computation of the modular Berry curvature is the parallel transportation equation. We will start by changing the modular Hamiltonian by acting with an element in the symmetry group $X\in \mathfrak{so}(2,d)$:
\be \label{eq:variation}
\delta H_{\rm mod}-P_0(\delta H_{\rm mod})=[X,H_{\rm mod}]~.
\ee
These \emph{shape-changing} variations change the spherical region without modifying the global state of the CFT\footnote{Of course, as explained in Section~\ref{sec:stresstensordef}, one can equivalently think of them in terms of a procedure where we keep the subregion fixed, but change the global state by insertion of a stress tensor operator in the Euclidean path integral.}. 
Recall that \eqref{eq:variation} is a special example of \eqref{eq:deltaHmod} where we take the diffeomorphisms to be conformal Killing vectors. 

Clearly, not all generators $X$ in \eqref{eq:variation} lead to a change of the modular Hamiltonian. The ones which satisfy $\delta H_{\rm mod} =0$, are the modular zero modes, and are formally defined as elements $Q\in \mathfrak{so}(2,d)$ which commute with the modular Hamiltonian: 
\be 
[Q,H_{\rm mod}]=0~.
\ee
This is precisely the definition of the stabilizer subalgebra $\mathfrak{h}\equiv \mathrm{stab}(H_{\rm mod})$. In the case that $H_{\rm mod}$ is given by \eqref{eq:Hmod} a suitable basis for the space of zero modes can be given by
\begin{align}
Q_i&=\frac{1}{2R} \left[-(R^2 + |\vec{x}_0|^2)P_i - 2 x_{0i} D+2 x_0^jM_{ij}+2x_{0i} x_0^jP_j+C_i \right]~, \label{eq:zeromode1} \\
Q_{ij}&=M_{ij}+x_{0i}P_j-x_{0j}P_i~, \label{eq:zeromode2}
\end{align}
where $i,j=1,\ldots,d-1$. Indeed, using the conformal algebra one can explicitly check that
\be 
[Q_i,H_{\rm mod}]=[Q_{ij},H_{\rm mod}]=0~.
\ee
The first class of zero modes in \eqref{eq:zeromode1} correspond to `boosts' (directed in the $i$-th direction of the $\vec{x}$ plane) that preserve the causal diamond associated to the spherical region on the boundary. The  second class of zero modes, \eqref{eq:zeromode2},  rotate the spherical region while leaving the diamond invariant. The algebra of the zero modes is given by:
\be \label{eq:non-trivialcommutators}
[Q_{i},Q_{j}]=Q_{ij}~, \hspace{15pt} [Q_{i},Q_{jk}]=Q_{k}\delta_{ij}-Q_{j}\delta_{ik}~.
\ee
Together with the modular Hamiltonian itself, the zero mode space $\mathcal{A}_0$ can therefore be identified with the subalgebra
\be \label{eq:zeromodespace}
\mathfrak{h}=\mathfrak{so}(1,1)\times \mathfrak{so}(1,d-1)~.
\ee
Note that the space of zero modes has a non-abelian component $\mathfrak{so}(1,d-1)$. 

\begin{figure}[t!]
    \centering
    \includegraphics[width=0.55\linewidth]{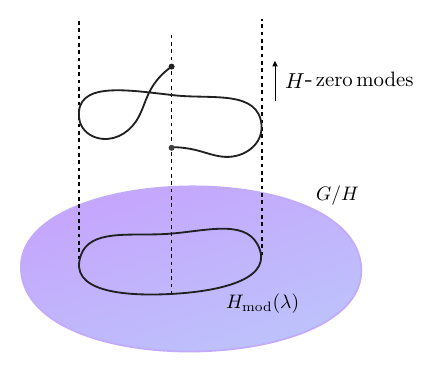}
    \caption{A parallel transport problem. The Berry curvature is associated to the principal $H$-bundle defined by $G\to G/H$ with fibers that are isomorphic to $H$. A closed curve of modular Hamiltonians $H_{\rm mod}(\lambda)$ in the base space $G/H$ is parallel lifted (using the Berry connection) to a non-closed curve in the group $G$. The endpoints of the curve differ by an element in the zero mode space $H$.}
    \label{fig:paralleltransport}
\end{figure}
	
The general structure of the modular Berry transport can now be described as follows: The space of modular Hamiltonians that we consider is locally given by the variations \eqref{eq:variation} and therefore parametrized by $X\in \mathfrak{g}/\mathfrak{h}$. 
Exponentiating, we conclude that the parameter space  is given by the coset space
\be \label{eq:coset}
\mathcal{O}_{H_{\rm mod}}\equiv \frac{SO(2,d)}{SO(1,1)\times SO(1,d-1)}~.
\ee
This is nothing other than the coset space describing the space of causal diamonds in a $d$-dimensional CFT, known as kinematic space~\cite{Czech:2015qta, Czech:2016xec, deBoer:2015kda, deBoer:2016pqk}.

The action of the symmetry group on parameter space is through conjugation and the subgroup of zero modes satisfies
\be \label{eq:zeromodeamb}
VH_{\rm mod}V^{-1}=H_{\rm mod}
\ee
for $V\in H$. A path in the coset space \eqref{eq:coset} can be identified with a one-parameter family of modular Hamiltonians. 
One can think of this as describing a fiber bundle\footnote{This defines a principal $H$-bundle in the following sense. There is an action of $H$ on $G$ through left-action:
\be
U\to VU~,
\ee
which is compatible with the projection $G\to G/H$, and the isomorphism $\mathfrak{g} \cong \mathfrak{g} /\mathfrak{h}\oplus \mathfrak{h}$ implies that the group $G$ is locally isomorphic to the trivial principal $H$-bundle.}
\be \label{eq:fiberbundle}
G\to G/H~,
\ee
which geometrizes the zero mode ambiguity in \eqref{eq:zeromodeamb} by associating to each modular Hamiltonian in the parameter space a fiber of zero modes that projects to the same element (see Figure \ref{fig:paralleltransport}). 

On an abstract level, the modular Berry connection now corresponds to a one-form on $G$ that takes values in the non-abelian zero mode space. Similarly, the Berry curvature $F$ takes the general form 
\be \label{eq:modBerrycurvature}
F=F^{H_{\rm mod}}H_{\rm mod}+\sum_{Q}F^{Q} Q~,
\ee
where the sum over $Q$ indicates a sum over a suitable basis of zero modes (excluding $H_{\rm mod}$ itself). Hence, $F\in \mathfrak{h}$ takes values in a non-abelian zero mode space, and satisfies $[H_{\rm mod}, F]=0$. One can compute the Berry curvature associated to two transformations $X_1,X_2$ from the general formula
\be 
F=P_0([X_1,X_2])~.
\ee
The map $P_0:\mathfrak{g}\to \mathfrak{h}$ denotes the zero mode projector, that extracts the component of the commutator in these directions. Explicitly, decomposing an arbitrary operator $X$ as
\be \label{eq:zeromodeprojector}
X=\alpha H_{\rm mod}+\sum_{Q}\alpha_Q Q+[H_{\rm mod},Y]~,
\ee
the projection operator will extract the parts with coefficients $\alpha$ and $\alpha_Q$.

Given the non-abelian structure of the zero mode space \eqref{eq:zeromodespace}, one needs to decompose the projector $P_0$ into subprojectors that extract each of the coefficients in \eqref{eq:zeromodeprojector} separately. In general, without introducing more structure, there is no unique procedure for doing this. In fact, one can simply redefine the operators $Q$ that constitute the zero mode basis to get a new set of coefficients $\alpha_Q$ in \eqref{eq:zeromodeprojector}. However, at this point we can use the fact that we are working with a finite-dimensional Lie algebra and introduce the notion of inner product $\langle \cdot, \cdot \rangle$ on the zero mode space. By choosing an orthonormal basis of zero modes, one can easily distinguish between them. A natural choice of inner product on the Lie algebra $\mathfrak{so}(2,d)$ is the Cartan-Killing form given by
\be \label{eq:Cartan-Killing}
\langle X, Y\rangle\equiv \frac{1}{2}\mathrm{tr}(XY)~,
\ee
where the trace is taken in the fundamental representation. Let us now choose a linearly independent set of zero mode generators $Q_a$ which are orthonormal with respect to the metric:
\be 
\langle Q_a, Q_b \rangle = \delta_{ab}~.
\ee 
Such an orthonormal basis can, for example, be obtained using the Gram-Schmidt procedure. Moreover, we require that  $\langle H_{\rm mod}, Q_{a}\rangle =0$. One can use the metric and corresponding orthonormal basis to extract the coefficients from the operator $X$. For example, we can define the projection $P_0^{H_{\rm mod}}$ on the $H_{\rm mod}$-component of the operator through
\be \label{eq:PHmod}
P^{H_{\rm mod}}_0(X)\equiv c_{H_{\rm mod}}^{-1}\langle H_{\rm mod}, X\rangle=\alpha~,
\ee
where the normalization is such that $c_{H_{\rm mod}}=\langle H_{\rm mod},H_{\rm mod}\rangle$. One can check that \eqref{eq:PHmod} indeed satisfies the properties that we usually associate with a projection
\be \label{eq:projcond}
P^{H_{\rm mod}}_0(H_{\rm mod})=1~, \hspace{15pt} P^{H_{\rm mod}}_0(Q_a)=0~, \hspace{15pt}  P^{H_{\rm mod}}_0([H_{\rm mod},Y])=0~,
\ee
by using the orthogonality of zero modes. Moreover, the last equality in \eqref{eq:projcond} can be proved using the cyclicity of the trace 
\be 
\mathrm{tr}(H_{\rm mod}[H_{\rm mod},Y])=0~.
\ee
Using this explicit form of the subprojector, we can compute the curvature component of \eqref{eq:modBerrycurvature} in the direction $H_{\rm mod}$ via the formula
\be \label{eq:Fmod}
F^{H_{\rm mod}}=P_0^{H_{\rm mod}}([X_1,X_2])~.
\ee

The non-abelian part of the curvature $F$ can be extracted in a similar fashion. To this end, we construct the subprojection operators onto the other zero modes
\be 
P_0^{Q_a}(X)\equiv c_{Q_a}^{-1}\langle Q_a,X \rangle=\alpha_{Q_a}~, 
\ee
and a different normalization $c_{Q_a}=\langle Q_a,Q_a \rangle$. In particular, the curvature component in the $Q_a$-direction is given by $F^{Q_a}=P_0^{Q_a}([X_1,X_2])$. This gives a concrete prescription for computing all the components of the modular Berry curvature in the case of shape-changing transformations. We will now show that the numbers that we extract from the operator $F$ can be computed from a symplectic form on certain coadjoint orbits of the conformal group. 

\subsubsection{Relation to coadjoint orbits}
Recall that the parameter space \eqref{eq:coset} of the modular Hamiltonian associated to shape-changing transformations is given by
\be \label{eq:kinematicspace}
\mathcal{O}_{H_{\rm mod}}=\frac{SO(2,d)}{SO(1,1)\times SO(1,d-1)}~.
\ee 
We will now observe that this has the structure of a geometry known as a coadjoint orbit.

Consider our algebra $\mathfrak{g} = \mathfrak{so}(2,d)$. It admits a bilinear pairing $\langle \cdot,\cdot \rangle$ given by~\eqref{eq:bilinearpairing} between elements of $\mathfrak{so}(2,d)$. Since the pairing is non-degenerate, the algebra and dual space $\mathfrak{g}^*$ (the space of linear maps on the algebra) are isomorphic. A coadjoint orbit is properly defined in terms of an orbit through the dual space, but due to this isomorphism it suffices to consider orbits of the algebra under a particular action: the adjoint action given by the Lie commutator. Such orbits form symplectic manifolds, and admit a symplectic form known as the Kirillov-Kostant symplectic form~\cite{kirillov2004, Witten:1987ty}. 

To define the Kirillov-Kostant symplectic form, let us first consider the Maurer-Cartan form
\be \label{eq:THETA}
\Theta = U^{-1} dU~,
\ee
on the group $U\in SO(2,d)$. Using the dual pairing 
the Kirillov-Kostant symplectic form is defined as
\be 
\omega \equiv \langle H_{\rm mod}, \Theta \wedge \Theta \rangle~.
\ee
To show that $\omega$ defines a symplectic form we use the Maurer-Cartan equation:
\be \label{eq:Maurer-Cartan}
d\Theta + \Theta \wedge \Theta =0~.
\ee
Indeed, from \eqref{eq:Maurer-Cartan} it immediately follows that $d(\Theta \wedge \Theta)=0$ which shows that $d\omega=0$. Hence, $\omega$ indeed defines a closed form on the group. Moreover, one can check from the definition \eqref{eq:THETA} that 
\be 
\Theta \wedge \Theta (X_1,X_2)=[X_1,X_2]~,
\ee
so that the Kirillov-Kostant form can also be written as
\be \label{eq:KKsymp}
\omega(X_1,X_2)=\langle H_{\rm mod}, [X_1,X_2] \rangle~.
\ee

Due to the presence of zero modes, \eqref{eq:KKsymp} has degeneracies when defined on the full group. The fact that $\omega$ descends to a symplectic form on the parameter space $\mathcal{O}_{H_{\rm mod}}$ follows from the observation:
\be 
\mathrm{tr}(H_{\rm mod}[X_1,X_2])=-\mathrm{tr}([X_1,H_{\rm mod}]X_2)=0~,
\ee
whenever $X_1 \in \mathrm{stab}(H_{\rm mod})$. Because the stabilizer of the modular Hamiltonian is precisely given by the subgroup $H=SO(1,1)\times SO(1,d-1)$, this shows that $\omega$ defines a symplectic form on the coadjoint orbit. Note that \eqref{eq:KKsymp} agrees with the formula \eqref{eq:Fmod} for $F^{H_{\rm mod}}$ up to a normalization constant. Thus, we find that the abelian part of the modular Berry curvature equals the Kirillov-Kostant symplectic form on kinematic space. This result was anticipated for the case $d=2$ in \cite{deBoer:2021zlm}, and we have now established it here in full generality. 
	
For the non-abelian part of the curvature, the situation is slightly different, in the sense that 
\be 
F^{Q_a}=0
\ee
on the stabilizer $\mathrm{stab}(Q_a)$, which consists of elements that commute with the zero mode $Q_a$. Of course, $H_{\rm mod}$ is such a stabilizing element (by definition of $Q_a$), but in general $\mathrm{stab}(H_{\rm mod})\neq \mathrm{stab}(Q_i)$. Therefore, the non-abelian components of the curvature do not descend a two-form on $\mathcal{O}_{H_{\rm mod}}$, but on a different coadjoint orbit $\mathcal{O}_{Q_a}$. Of course, this coadjoint orbit has the same global structure as \eqref{eq:kinematicspace} (from a mathematical perspective there is nothing special about the zero mode $H_{\rm mod}$ compared to the other $Q_a$), but the explicit parametrization in terms of conformal group generators will be different. The rest of the arguments that were given above still go through, so that we can identify the $Q_{a}$-component of the shape-changing Berry curvature with the Kirillov-Kostant symplectic form on $\mathcal{O}_{Q_a}$. 
    
\subsubsection{Low-dimensional examples}
Let us now work out some low-dimensional examples, and use the parallel transport formalism to compute the modular Berry curvature by changing the shape of the entangling region. The results will agree with the Crofton formula for computing lengths of geodesics in the bulk~\cite{Czech:2015qta}. (For the higher-dimensional case, see~\cite{Balasubramanian:2018uus}.)  

\begin{figure}[t!]
	\centering
	\includegraphics[width=0.8\linewidth]{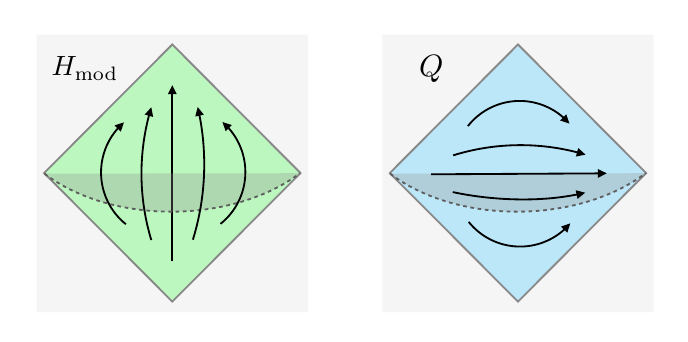} 
	\caption{Left: The action of $H_{\rm mod}$ on the causal diamond. Right: The action of $Q$ on the causal diamond. 
 }
	\label{fig:hmodvectorplot}
\end{figure}

We first restrict to the case of a CFT$_2$ on the plane. The entangling region on the boundary is an interval (specified by its midpoint $x_0$ and radius $R$), with modular Hamiltonian \eqref{eq:Hmod} given by
\be \label{eq:Hmod2}
H_{\rm mod} = \frac{\pi}{R} \left[-(R^2 - x_0^2 )P_0 - 2 x_0 M_{01}-C_0 \right]~.
\ee
For the unit-interval centered at the origin $(x_0,R)=(0,1)$ this expression reduces to $H_{\rm mod} = -\pi\left[P_0+C_0 \right].$ A representation of the corresponding vector field is provided in Figure \ref{fig:hmodvectorplot} (left panel). Note that it preserves the causal diamond associated to the interval. The case $d=2$ allows for one additional zero mode in \eqref{eq:zeromode1},  which we denote by $Q$, and is given by
\be 
Q=\frac{1}{2R} \left[-(R^2 -x_0^2)P_1 - 2 x_0 D+C_1 \right]~.
\ee
Again, for $(x_0,R)=(0,1)$ we have $Q= \frac{1}{2}\left[-P_1+C_1 \right]$. It amounts to a spatial `boost' that fixes the entangling surface at $x=x_0\pm R$, see Figure \ref{fig:hmodvectorplot} (right panel).

The Berry transport problem involves a modification of the entangling region by some generators of the conformal group. Given our parametrization of the modular Hamiltonian \eqref{eq:Hmod2} in terms of the midpoint $x_0$ and radius $R$ a natural choice of shape-changing transformations are translations and widenings of the interval. In these cases, the parallel transport equation in \eqref{eq:variation} becomes
\be \label{eq:berrytransport}
\partial_{x_0} H_{\rm mod} =\left[\mathcal{S}_{\delta x_0},H_{\rm mod}\right]~, \hspace{15pt} \partial_R H_{\rm mod} =\left[\mathcal{S}_{\delta R},H_{\rm mod}\right]~.
\ee
The operators that implement the changes in shape are denote by $\mathcal{S}_{\delta x_0}$ and $\mathcal{S}_{\delta R}$ respectively. 
Using the commutation relations in Appendix \ref{appendixA}, 
it is easy to see that the parallel transport operator for translations is
\be 
\mathcal{S}_{\delta x_0}=P_{1}~.
\ee
Similarly, one can show that 
\be 
\mathcal{S}_{\delta R}=-\frac{1}{R}\left(x_0P_1-D\right)~.
\ee
Let us now study the relevant subprojection operators $P_0^{H_{\rm mod}}$ and $P_0^{Q}$. The modular Berry curvature associated to this parallel transport problem is given by 
\be \label{eq:F2}
F^{H_{\rm mod}}=0~,\hspace{20pt} F^{Q}=P_0^{Q}([\mathcal{S}_{\delta x_0},\mathcal{S}_{\delta R}])=\frac{1}{R^2}~.
\ee
Note that it is proportional to the zero mode $Q$, and for this reason naturally lives on the kinematic space of boundary intervals in CFT$_2$:
\be
\mathcal{O}_Q=\frac{SO(2,2)}{SO(1,1)\times SO(1,1)}~.
\ee
The associated $(x_0,R)$-component of the Kirillov-Kostant symplectic form is now given by
\be \label{eq:sympformQ}
\omega_{Q} = \frac{1}{R^2} d x_0 \wedge dR~.
\ee

We can rewrite \eqref{eq:sympformQ} in a more familiar form by using a cylindrical coordinate $\theta$ on the boundary time slice via the identification $x=\tan{(\theta/2)}$. In particular, identifying
\be 
R =\tan{(\alpha/2)}~,
\ee
where the parameter $\alpha$ measures the opening angle of the boundary subregion, the symplectic form \eqref{eq:sympformQ} at $x_0=0$ becomes 
\be 
\omega_Q= \frac{1}{4\sin^2(\alpha/2)}d\theta\wedge d\alpha~.
\ee
 This result agrees with the well-known Crofton formula for RT surfaces on the hyperbolic disk~\cite{Czech:2015qta}, which is identified with the $t=0$ time slice of $\mathrm{AdS}_3$. In particular, it can be used to compute lengths of curves in the bulk.

Note that the full symplectic form on $\mathcal{O}_Q$ also includes information about shape-changes that, for example, tilt the interval, and take it away from the fixed time slice. To access this information one would need to compute the components of the curvature associated to these deformations as well. For now we will restrict to changes implemented by $\mathcal{S}_{\delta x_0}$ and $\mathcal{S}_{\delta R}$ as in \eqref{eq:F2}, that act within a single time slice.

Let us also consider the case of CFT$_3$, where we take the boundary region to be a disk on the $(x^1,x^2)$-plane with radius $R$. According to \eqref{eq:Hmod} the modular Hamiltonian associated to this spherical region is given by:
\be 
H_{\rm mod} = \frac{\pi}{R} \left[-(R^2 - (x_0^1)^2-(x_0^2)^2)P_0 - 2(x_0^1 M_{01}+x_0^2 M_{02})-C_0 \right]~.
\ee
For the unit-circle this again reduces to the simple expression $H_{\rm mod} = -\pi\left[P_0+C_0 \right]$. There are three distinct zero modes, as can be seen from \eqref{eq:zeromode1} and \eqref{eq:zeromode2}:
\begin{align}
Q_1&=\frac{1}{2R} \left[-(R^2 + (x_0^2)^2-(x_0^1)^2)P_1 - 2 x_0^1 D+2 x_0^2M_{12}+2x_0^1 x_0^2P_2+C_1 \right]~,\\
Q_2&=\frac{1}{2R} \left[-(R^2 + (x_0^1)^2-(x_0^2)^2)P_2 - 2 x_0^2 D-2 x_0^1M_{12}+2x_0^2 x_0^1P_1+C_2 \right]~,\\
Q_3&\equiv Q_{12} = M_{12}+x_0^1P_2-x_0^2P_1~,
\end{align}
which constitute a non-abelian $\mathfrak{so}(1,2)$ algebra:
\be 
[Q_1,Q_2]=Q_3~, \hspace{15pt} [Q_1,Q_3]=Q_2~,  \hspace{15pt} [Q_2,Q_3]=-Q_1~.
\ee
These correspond to two `spatial' boosts and one rotation that preserve the spherical entangling region $|\vec{x}-\vec{x_0}|=R$. 
The Berry transport equations \eqref{eq:berrytransport} are unchanged, except that we now have two translations indicated by $\mathcal{S}_{\delta x_0^j}$, with $j=1,2$. These are given by: 
\be 
\mathcal{S}_{\delta x_0^1}=P_{1}~,\hspace{15pt} \mathcal{S}_{\delta x_0^2}=P_{2}~,\hspace{15pt} \mathcal{S}_{\delta R}=\frac{1}{R}\left(D-x_0^1P_1-x_0^2P_2\right)~.
\ee
Now we can compute the commutator associated to a change of center position and a change of radius to be
\beq 
[\mathcal{S}_{\delta x_0^j},\mathcal{S}_{\delta R}]=-\frac{1}{R}P_j~.
\eeq
Hence, we find that the component of the Berry curvature in the $H_{\rm mod}$-direction vanishes:
\beq 
F^{H_{\rm mod}}=P^{H_{\rm mod}}_0\left([\mathcal{S}_{\delta x_0^j},\mathcal{S}_{\delta R}]\right)= -\frac{1}{R}P^{H_{\rm mod}}_0(P_j)=0~.
\eeq
The component in the $Q_i$-direction will be non-zero. Indeed, the curvature is given by
\beq 
F^{Q_i}=-\frac{1}{R}P_0^{Q_i}(P_j)=\delta_{ij}\frac{1}{R^2}~, \hspace{10pt} \mathrm{for} \hspace{10pt} i=1,2~,
\eeq
and $F^{Q_3}=0$. 
Note that the non-zero component of the curvature depends on the direction of the translation: Acting with $\mathcal{S}_{\delta x_0^j}$ leads to $F^{Q_j}\neq 0$. Similarly by setting the center to $\vec{x}_0=0$, the relevant component of the symplectic form in the $(x_0^i,R)$-direction is again given by: 
\beq 
\omega_{Q_i} = \frac{1}{R^2} d x_0^i \wedge dR~.
\eeq

\section{Discussion}

We have considered modular parallel transport involving a change of \emph{state} in holography in general dimensions. The resulting modular Berry curvature, which is operator-valued, contains information about both the bulk symplectic form as well as the quantum Fisher information metric and its bulk dual, the canonical energy. We additionally treated shape-changing modular transport in higher dimensions, which is a special case of the state-changing transformations, and in this case provided a connection to the geometry of coadjoint orbits.

One could interpret the current work as a continuation of \cite{deBoer:2021zlm}, where the modular Berry phase is studied in the specific example of AdS$_3$/CFT$_2$,  extended to a larger class of state deformations and to the higher-dimensional setting. Of course, that setting is rather special in the sense that certain properties of AdS$_3$ gravity and two-dimensional CFTs do not generalize to higher dimensions. For example, the state-changing transformations that were considered in \cite{deBoer:2021zlm} are \emph{not} part of the symmetry algebra of $\mathrm{CFT}_d$ when $d>2$. In higher dimensions, the finite-dimensional conformal group only contains shape-changing transformations. This is to be contrasted with $\mathrm{CFT}_2$, where we have the full infinite-dimensional Virasoro algebra at our disposal. To set up a non-trivial transport problem in higher dimensions we had to introduce a more general formalism of coherent state deformations, that are not restricted to act within the symmetry algebra. Another important difference arises in the bulk computation: While AdS$_3$ has a topological Chern-Simons theory description that makes the computation of the symplectic form somewhat tractable, no such simplification happens in general Einstein-Hilbert gravity. In the present work, we instead use the covariant phase space formalism directly in the metric formalism to find an expression for the bulk symplectic form. However, as we have shown here, the relation between the Berry phase and symplectic form persists even in this more general setting.  

We should also discuss our results in light of previous work on the role of Berry phases in the AdS/CFT correspondence. A notable example involves \cite{Belin:2018bpg,Belin:2018fxe} where an interesting connection between the Berry phase and bulk symplectic form is established. Their computation involves the space of coherent \emph{pure} states that are prepared via the Euclidean path integral by turning on sources. The corresponding Berry phase is shown to agree with the bulk symplectic form associated to the full Cauchy slice. Our approach involves a similar set-up with the important difference that our computations work for deformations of density matrices associated to general subregions in the CFT. The corresponding bulk dual is now the symplectic form supported on the entanglement wedge. In that sense, our work provides a natural extension of these previous results to CFT subregions, and places the Berry phase/bulk symplectic form duality on a more general footing. Another approach is to consider other geometric phases in AdS/CFT and their relations to wormholes. This has been recently studied in \cite{Nogueira:2021ngh, Banerjee:2022jnv}.

To associate a geometric phase to deformations of density matrices we used the construction of the \emph{modular} Berry phase. It is built upon the idea that there is a zero mode ambiguity in the choice of basis frame for the modular Hamiltonian. There is a slightly different version of the parallel transport problem due to Uhlmann that relies on the idea of parallel purifications \cite{Uhlmann:1976def,Uhlmann:1990er}. The resulting Uhlmann holonomy is closely related to, but not exactly the same as the modular Berry curvature. One difference is that the Uhlmann equations are written in terms of the change of density matrix itself while the modular Berry curvature makes use of the change of the modular Hamiltonian as a starting point. There is a non-trivial transformation, cf. \eqref{eq:deltaHdeltarho}, that relates both perspectives. More importantly, the zero mode projection that is crucial in defining modular Berry transport is absent in the Uhlmann case. While the Uhlmann holonomy is also related to a distance measure on the space of mixed states, i.e., the fidelity, our results indicate that the modular Berry phase is instead related to the quantum Fisher information metric on the space of mixed states. To understand this more deeply would be useful for many reasons. For example, the Uhlmann holonomy was used by \cite{Kirklin:2019ror} to make a claim that is similar in spirit to ours: that there is a direct connection between the geometric phase and some bulk entanglement wedge symplectic form. 

The fact that the metric and the symplectic form are related in a simple way through \eqref{Gbulk} suggests an underlying geometric structure. In fact, the relation immediately brings to mind the situation for a K\"ahler manifold where the symplectic form and metric are related by an extra insertion of the (almost) complex structure. This is familiar from the usual Berry phase in finite-dimensional quantum mechanics, where the space of pure states takes the form of a complex projective space, which does indeed exhibit a natural K\"ahler structure. It is well known that in this case the Berry curvature is closely related to the Fubini-Study metric. However, in the case of mixed states we have found that to go from the modular Berry curvature to the quantum Fisher information metric requires an extra action of the modular Hamiltonian. This procedure does not seem to have a natural interpretation as an almost complex structure: Importantly, it does not square to minus one when acting on general tangent vectors. Only in special cases (for example, when we are acting purely with shape-changing transformations) do we expect that the presence of such underlying geometric structure can be made precise. Nevertheless it would be interesting to understand these observations better. 

Likewise, one might ask whether this generalized symplectic structure defines a natural Hilbert space through geometric quantization. In the the shape-changing case, recall that the Berry curvature could be related to the Kirillov-Kostant symplectic form on a special symplectic geometry known as a coadjoint orbit. By the `orbit method,' which is a version of geometric quantization, such symplectic manifolds can be equated with a particular representation of the group which defines the coadjoint orbit through quotienting~\cite{kirillov2004}. In this more general setting involving state-changes, it would be  interesting to learn if similar relations persist, and what one can learn from them about the Hilbert space for quantum gravity.

Recently, the role of operator algebra techniques has gained some renewed interest in the context of holography and black hole physics. In particular, it was argued that the large $N$ limit of the boundary CFT (in the specific setting of the eternal black hole) should be a type III$_1$ von Neumann algebra \cite{Leutheusser:2021qhd,Leutheusser:2021frk}. Type III von Neumann algebras are rather complicated in the sense that many quantities that we like to use in quantum mechanics (e.g., density matrices, von Neumann entropies) are not well-defined. It is therefore natural to ask how our computations depend on details of the underlying operator algebra. Crucially, entropy differences (e.g., the relative entropy) are well-defined in type III von Neumann algebras. Since the final answer for the Berry curvature is related to the quantum Fisher information metric, which can be written in terms of the relative entropy, it is certainly possible that there exists some suitable continuum limit of our computations. One idea is to define a version of the Berry phase problem in terms of the algebra of observables without any reference to an underlying state deformation. It would be interesting to study further the Berry phase in connection with the emergent type III$_1$ structure. 

A related question involves the possibility of including spectral deformations in the parallel transport problem.\footnote{We thank Erik Verlinde for making this suggestion.} Na{\"\i}vely, in our setup there does not seem to be any non-trivial contribution to the curvature from transformations that act purely within the diagonal part of the modular Hamiltonian. This can be seen most easily from the parallel transport equation \eqref{paralleltransport}. If we diagonalize the modular Hamiltonian via some unitary $U$ it follows that  
\beq 
H_{\rm mod}=U\Delta U^{\dagger}~, \hspace{15pt} P_0(\delta H_{\rm mod})= U \delta \Delta U^{\dagger}~.
\eeq
The zero mode piece of $\delta H_{\rm mod}$ therefore consists of the changes in the modular spectrum, and it is precisely this term that is subtracted in defining the modular transport operator. Perhaps it is possible to modify the setup in such a way as to keep track of the spectral deformations as well, and include their effect in some generalized curvature. 

\section*{Acknowledgements}
It is a pleasure to thank Tom Faulkner, Alex May and Erik Verlinde for discussions. BC and RE are supported by the Dushi Zhuanxiang Fellowship. RE also acknowledges a Shuimu Scholarship as part of the “Shuimu Tsinghua Scholar” Program. RE was supported by the ERC Consolidator Grant QUANTIVIOL. BN is supported by the Spinoza Grant of the Dutch Science Organisation (NWO). JdB and JvdH are supported by the European Research Council under the European Unions Seventh Framework Programme (FP7/2007-2013), ERC Grant agreement ADG 834878. CZ is supported by a UMD Higholt Professorship, and acknowledges a Heising-Simons Fellowship as part of the “Observational Signatures of Quantum Gravity” collaboration grant 2021-2818. This work is supported by the
Delta ITP consortium, a program of the Netherlands Organisation for Scientific Research (NWO) that is funded by the Dutch Ministry of Education, Culture and Science (OCW).

\appendix
\section{Conformal algebra} \label{appendixA}
We will review here some facts about the $d$-dimensional conformal algebra, which will set our conventions throughout the paper. 

The conformal generators are
\begin{align}
    D &= - x^\mu \partial_\mu~, \hspace{10pt}P_\mu = -\partial_\mu~, \hspace{10pt}C_\mu = x^2\partial_\mu - 2x_\mu x^\rho \partial_\rho~,\hspace{10pt}M_{\mu \nu} = x_\mu \partial_\nu - x_\nu \partial_\mu~.
\end{align}
The resulting commutation relations are given by
\begin{align}
\left[D,P_\mu \right] & = P_\mu~,  \hspace{20pt}\left[ D, C_\mu \right] = - C_\mu~, \nonumber \\ \left[C_\mu , P_\nu \right]& =2 (\eta_{\mu \nu} D - M_{\mu \nu})~, 
\nonumber \\
  \left[  M_{\mu \nu}, P_\rho \right] &=- \eta_{\mu \rho} P_\nu+\eta_{\nu \rho} P_\mu ~, \nonumber \\
  \left[ M_{\mu \nu}, C_\rho \right] &= - \eta_{\mu \rho} C_\nu+\eta_{\nu \rho } C_\mu~, ~ \nonumber \\  \left[ M_{\mu \nu} ,M_{\sigma\rho}\right] &= -\eta_{\mu \sigma}M_{\nu \rho} + \eta_{\nu \sigma} M_{\mu \rho} - \eta_{\nu \rho}M_{\mu \sigma}  + \eta_{\mu \rho}M_{\nu \sigma}~.
\end{align}
Note that we have written $\mu = (0,i)$, where $i=1,\ldots, d-1$ is some spatial index.

The bilinear product on the conformal algebra is given by
\begin{equation}\label{eq:bilinearpairing}
\langle X,Y \rangle \equiv \frac{1}{2}\mathrm{tr}(XY)~, \hspace{10pt} X,Y\in \mathfrak{so}(2,d)~,
\end{equation}
where the trace is taken in the fundamental representation. In terms of the above generators the inner product is normalized such that non-zero entries are given by:
\begin{align}
\langle D,D \rangle = \langle M_{0i},M_{0i} \rangle = - \langle M_{ij},M_{ij} \rangle = 1~, \hspace{10pt}  \langle P_{0},C_{0} \rangle = - \langle P_{i},C_{i} \rangle = 2~.
\end{align}

\section{Relative entropy and quantum Fisher information} 
\label{sec:relativeentropy}
In this appendix we review the derivation of a metric on the space of density matrices from the second variation of the relative entropy \cite{Lashkari:2015hha,Faulkner:2017tkh}. The relative entropy between two states $\sigma$ and $\rho$ is given by:
\beq 
S(\sigma||\rho)\equiv \mathrm{tr}(\sigma\log\sigma)-\mathrm{tr}(\sigma\log\rho)~.
\eeq
Let us view $\sigma$ as obtained from $\rho$ by some small perturbation: 
\beq
\sigma = \rho+\varepsilon \delta \rho + O(\varepsilon^2)~,
\eeq
where $\varepsilon$ is some small parameter. Then, the second derivative with respect to this parameter can be expressed as:
\beq
\frac{d^2}{d\varepsilon^2} S(\sigma||\rho) = \mathrm{tr}\left(\delta \rho \frac{d}{d\varepsilon}\log(\rho+\varepsilon\delta \rho)\right)~. \eeq
To compute the derivative we use the following integral representation for the logarithm of an operator:
\beq \label{eq:logformula}
\log(\rho+\varepsilon\delta \rho)= -\int_{0}^{\infty} \frac{ds}{s}\left(e^{-s(\rho+\varepsilon\delta \rho)}-e^{-s}\right)~.
\eeq
One can now take the derivative by using the relation
\beq 
\frac{d}{d\varepsilon} e^{A+\varepsilon B} = \int_0^1 dx \, e^{Ax}Be^{(1-x)A}~,
\eeq
for two operators $A$ and $B$. Using \eqref{eq:logformula} it now follows that
\beq 
\frac{d^2}{d\varepsilon^2} S(\sigma||\rho) = \int_0^1 dx\int_0^{\infty} ds \, \mathrm{tr}\left(\delta \rho\, e^{-x s\rho}\delta \rho \,e^{-(1-x)s\rho}\right)~.
\eeq
We can now evaluate the trace in the eigenbasis of the modular Hamiltonian associated to the state $\rho$:
\beq 
\rho|\omega \rangle = e^{-\omega}|\omega\rangle~.
\eeq
We can write this as:
\begin{align}
\frac{d^2}{d\varepsilon^2} S(\sigma ||\rho) &= \int_0^1 dx\int_0^{\infty} ds \, \int d\omega \int d\omega' |\langle \omega |\delta \rho |\omega'\rangle|^2 e^{-sx(e^{-\omega}-e^{-\omega'})} e^{-se^{-\omega'}} \nonumber \\
&=\int d\omega \int d\omega'|\langle \omega |\delta \rho |\omega'\rangle|^2e^{\omega}(\omega-\omega')n(\omega-\omega')~.
\end{align}
Using again the sinh-formula \eqref{eq:sinhintegral} to replace the integral over frequencies by an integral over modular time, and removing the explicit $|\omega\rangle$ basis we find that this expression is equivalent to
\begin{align} \label{eq:secondvariation}
\delta^{(2)} S(\sigma||\rho) &= \int_{-\infty -i\epsilon}^{\infty -i \epsilon}ds \frac{\pi}{2\sinh^2(\pi s)} \mathrm{tr}(\rho^{-1} \delta \rho \,\rho^{-is} \delta \rho \,\rho^{is})~.
\end{align}
This is an expression for the second-order variation of the relative entropy. We will now define a metric on the space of quantum states starting from the above expression. The second derivative of the relative entropy \eqref{eq:secondvariation} is a quadratic function in the state perturbations $\delta \rho$, so we can upgrade it to a bilinear form by taking two (possibly) different variations $\delta_1\rho,\delta_2\rho$ on the right-hand side. Plugging in the expressions for $\delta \rho$ in terms of the operators $\mathcal{O}$ using \eqref{eq:densitymatrixchange} we find that
\begin{align}
\delta_1\delta_2 S(\sigma ||\rho) = \int d^dx \int d^dx'\, \delta_1 \lambda(x) \delta_2\lambda(x') \int_{-\infty -i\epsilon}^{\infty-i\epsilon}ds \frac{\pi}{2\sinh^2(\pi s)}\langle \mathcal{O}(x) \mathcal{O}_s(x')\rangle~,
\end{align}
where the expectation value is taken in the reference state $\rho$. This is also known as the \emph{quantum Fisher information metric} \cite{Lashkari:2015hha,Faulkner:2017tkh}. 
This expression agrees with the `metric' $G_{\Psi}$ associated to the modular Berry curvature \eqref{eq:modularBerrymetric}. We have therefore established our identification.
\section{An explicit bulk computation}\label{sec:gravsympshape}

In this appendix we will give a short computation of the bulk dual of the shape-changing Berry transport problem. This is a special example of the computation that was done for more general diffeomorphisms in Section~\ref{sec:stresstensordef}. The first part of the computation closely follows  Section~3.2 of \cite{Kraus:2021cwf}. 

\subsection{Gravitational symplectic form for shape deformations}

We start from the gravitational symplectic form associated to the entanglement wedge:
\beq \label{eq:bulksympformfinal}
\Omega(\delta_1 g,\delta_2 g) = \int_{\Sigma} \omega~.
\eeq
We would like to evaluate this expression on metric perturbations that correspond to conformal Killing vector fields on the boundary, i.e., are in the conformal algebra $\mathfrak{so}(2,d)$. Given a solution $\xi$ to the conformal Killing equation 
\beq  \label{eq:CKV}
\partial_\mu \xi_\nu+\partial_\nu \xi_\mu=\frac{2}{d} h_{\mu \nu} \partial_\alpha \xi^\alpha~,
\eeq
one can construct a bulk vector field $\zeta$ that leaves the boundary metric invariant $\delta_{\zeta}h=0$. We assume that the asymptotic AdS$_{d+1}$ bulk metric $g$ is in Fefferman-Graham (FG) gauge so that we can write
\be \label{eq:asymptoticmetric1}
ds^2= \frac{d\rho^2}{4\rho^2}+\gamma_{\mu\nu}(\rho, x) dx^\mu dx^\nu~
\ee
near the asymptotic boundary $\rho\to 0$. The boundary metric $ds^2_h$ can be extracted from
\be \label{eq:asymptoticmetric2}
\gamma_{\mu\nu}(\rho,x)\sim \rho^{-1}h_{\mu\nu}(x)+\mathcal{O}(\rho^0)~, \hspace{15pt} ds^2_h =dt^2+dr^2+r^2d\Omega_{d-2}^2~.
\ee
The conformal factor that arises in the boundary metric $\delta h$ can be reabsorbed by introducing a non-trivial $\rho$-dependence in the vector field $\zeta$.
We use $\zeta$ to perturb the bulk metric. It turns out that $\omega$ in \eqref{eq:bulksympformfinal} becomes exact in spacetime, i.e., $i_{\zeta} \omega = d \mathcal{X}$, where $\mathcal{X}$ is a one-form on field space and a $(d-1)$-form on spacetime. There is an expression for $\chi$ of the form $\mathcal{X}=\chi\cdot \epsilon$, where \cite{Kraus:2021cwf}
\begin{align} \label{eq:chii}
\chi^{\mu \nu}=&\frac{1}{16\pi G}\left(\left(\nabla_{\alpha}\delta g^{\alpha \nu}+\nabla^{\nu}\delta \ln g \right)\xi^{\mu}+\nabla^{\nu} \delta g^{\mu \alpha}\xi_{\alpha}-\nabla_{\alpha}\xi^{\mu}\delta g^{\alpha \nu}-\frac{1}{2}\nabla^{\nu}\xi^{\mu}\delta \ln g \right)-(\mu \leftrightarrow \nu)~,
\end{align}
and $\delta \ln g \equiv g^{\mu\nu}\delta g_{\mu\nu}$. Using Stokes' theorem we can now write the symplectic form as a boundary integral
\beq 
i_{\xi}\Omega =  \int_{\Sigma} d\mathcal{X} =  \int_{
\partial \Sigma} \mathcal{X}~,
\eeq
over $\partial \Sigma$. The boundary of the entanglement wedge consists of two components (see Figure \ref{fig:entanglementwedge}): the asymptotic boundary region $A$ and the RT surface $\gamma_A$
\beq
\partial \Sigma = \gamma_A \cup A~.
\eeq
Following the general discussion, we will only consider the contribution coming from the asymptotic boundary. Given the geometry in \eqref{eq:asymptoticmetric1}, the symplectic form associated to the entanglement wedge now becomes 
\be \label{eq:bdysympform2}
i_{\xi}\Omega = \int_{A}d^{d-1}x \,\sqrt{g}\, \chi^{\rho t}~,
\ee
where the integration is over subregion $A$ at $\rho=t=0$, and $\chi^{\rho t}$ is a component of \eqref{eq:chii}. We can further simplify the expression for $\chi^{\rho t}$. The covariant derivative only acts non-trivially on $\delta g$ in the $\rho$-direction: we can replace $\nabla^{\mu}=4\rho^2 \delta_{\mu \rho}\partial_{\rho}$ when acting on $\delta g$ at $\rho=0$. Moreover, we can also use that $\partial_{\rho} \delta g_{\mu \rho}|_{\rho=0}=\partial_{\rho} \delta g^{\mu \rho}|_{\rho=0}=0$, since these components are fixed by the asymptotic form of the metric \eqref{eq:asymptoticmetric1}. Then, the component $\chi^{\rho t}$ simplifies to 
\be \label{eq:bdysymp2}
\chi^{\rho t}=-\frac{1}{16\pi G}\left(\nabla^{\rho}\delta \ln g \xi^{t}+\nabla^{\rho}\delta g^{\mu t}\xi_{\mu}\right)=-\frac{\rho^2}{4\pi G}\left(g^{\mu\nu}\partial_{\rho}\delta g_{\mu\nu} \xi^{t}-\partial_{\rho}\delta g_{\mu\nu}g^{\nu t}\xi^{\mu}\right)~,
\ee
where we have used that $\delta g^{\mu\nu} = -g^{\mu\lambda}g^{\nu \sigma}\delta g_{\lambda \sigma}$. We can rewrite this in terms of the expectation value of the stress tensor at the boundary  
\be
\mathcal{T}_{\mu\nu}\equiv \frac{1}{4 G}\left(K_{\mu\nu}-Kg_{\mu\nu}\right)~,
\ee 
where the extrinsic curvature is given by $K_{\mu\nu}=-\rho \partial_{\rho} g_{\mu\nu}$ and $K=g^{
\mu\nu}K_{\mu\nu}$. The holographic relation is given by $\mathcal{T}_{\mu\nu}=\langle \Psi| T_{\mu\nu} |\Psi \rangle$. One finds that\footnote{One can use the following variations
\be
\delta K_{\mu\nu}=-\rho \partial_{\rho} \delta g_{\mu\nu}~, \hspace{10pt} \delta K =- \rho g^{\mu\nu} \partial_{\rho}\delta g_{\mu\nu}~.
\ee}
\be 
\delta \mathcal{T}_{t\nu}=\frac{1}{4 G}\left(-\rho \partial_{\rho} \delta g_{t \nu}+ \rho g^{\mu \sigma} \partial_{\rho}\delta g_{\mu \sigma} g_{t\nu}\right)~.
\ee
This can be used to rewrite \eqref{eq:bdysymp2} as \be \label{eq:derivation1}
\chi^{\rho t}=-\frac{\rho^2}{4\pi G} \gamma^{tt}\left(g^{\mu \sigma}\partial_{\rho}\delta g_{\mu \sigma} \xi_{t}-\partial_{\rho}\delta g_{\nu t}\xi^{\nu}\right)=-\frac{\rho}{\pi} \sqrt{\gamma^{tt}}\delta T_{t \nu}n^{t}\xi^{\nu}~.
\ee
Here, we have introduced the normal vector $n$ to the $t=0$ time slice with component $n^{t}=\sqrt{\gamma^{tt}}$. Decomposing the metric \eqref{eq:asymptoticmetric1} according to $\sqrt{g}=(2\rho)^{-1}\sqrt{\gamma}=(2\rho)^{-1}\sqrt{\gamma_{tt}}\sqrt{\gamma^{(d-1)}}$, where the $(d-1)$-dimensional metric $\gamma^{(d-1)}$ is defined on the $t=0$ time slice which contains the spatial components of $\gamma$. Combining this with \eqref{eq:derivation1}, we obtain the expression: 
\beq
\omega\left(\delta_{\zeta_1} g, \delta_{\zeta_2} g\right)=-\frac{1}{2\pi}\int_A d S^\mu \,\xi^{\nu}_{1} \delta_{\xi_2} \mathcal{T}_{\mu \nu}~, \hspace{10pt} \mathrm{where} \hspace{10pt} d S^\mu = n^{\mu} \sqrt{\gamma^{(d-1)}} \, d^{d-1}x~, 
\eeq
in terms of the variation of the stress tensor profile $\delta_{\xi} \mathcal{T}_{\mu\nu}$. To further simplify the expression we use that for $\xi$ is a conformal Killing vector one has the following stress tensor transformation law\footnote{This formula holds for $d\geq 3$. In the case of two dimensions there is also the possibility of a central charge term. For the derivation, see Appendix~\ref{sec:Ttransform}.}:
\beq \label{eq:stresstransf}
\delta_{\xi} \mathcal{T}^{\mu \nu}=\xi^{\sigma} \partial_\sigma \mathcal{T}^{\mu \nu}+\left(\partial_\sigma \xi^\sigma\right) \mathcal{T}^{\mu \nu}-\left(\partial_\sigma \xi^\mu\right) \mathcal{T}^{\sigma \nu}+\left(\partial^\nu \xi_\sigma\right) \mathcal{T}^{\mu \sigma}~.
\eeq
Before using \eqref{eq:stresstransf} we first rewrite the symplectic form as
\beq
\omega\left(\delta_{\zeta_1} g, \delta_{\zeta_2} g\right)=-\frac{1}{2\pi}\int_{\tilde{A}} d^d x\left(\partial_\mu \xi_{1 \nu}\right) \delta_{\xi_2} \mathcal{T}^{\mu \nu}~,
\eeq
where we have introduced a region $\tilde{A}$ in the boundary CFT with $\partial \tilde{A}=A$, and used subsequently Stokes' theorem 
and the conservation equation $\partial_\nu\left(\delta_{\xi} \mathcal{T}^{\mu \nu}\right)=0$. Plugging in \eqref{eq:stresstransf} we find that
\beq \label{eq:somerewriting}
\omega=-\frac{1}{2\pi}\left[\int_{\tilde{B}} d^d x\left(\partial_\mu \xi_{1 \nu}\right)\left(-\left(\partial_\sigma \xi_2^\mu\right) \mathcal{T}^{\sigma \nu}+\left(\partial^\nu \xi_{2 \sigma}\right) \mathcal{T}^{\mu \sigma}\right)+\int_{\tilde{B}} d^d x\left(\partial_\mu \xi_{1 \nu}\right) \partial_\sigma\left(\xi_2^\sigma \mathcal{T}^{\mu \nu}\right)\right]~.
\eeq
The integrand of the first integral in \eqref{eq:somerewriting} can be rewritten in terms of the commutator
\beq
\left[\xi_1, \xi_2\right]_\nu=\xi_{1 \sigma} \partial^\sigma \xi_{2 \nu}-\xi_{2 \sigma} \partial^\sigma \xi_{1 \nu}~,
\eeq
using the fact that
\begin{align}
\left(\partial_\mu \xi_{1 \nu}\right)&\left(-\left(\partial_\sigma \xi_2^\mu\right) \mathcal{T}^{\sigma \nu}+\left(\partial^\nu \xi_{2 \sigma}\right) \mathcal{T}^{\mu \sigma}\right) 
=-\partial_\mu\left(\left[\xi_1, \xi_2\right]_\nu\right) \mathcal{T}^{\mu \nu}~.
\end{align}
Here, we have relabeled indices and used that $\partial_\mu \partial^\sigma \xi_\nu \mathcal{T}^{\mu \nu}=0$ (see \eqref{eq:identityforstress} for a derivation). 
The second integral in \eqref{eq:somerewriting} is zero. To see this, one can write
\beq \label{eq:bulkcommutator}
\left(\partial_\mu \xi_{1 \nu}\right) \partial_\sigma\left(\xi_2^\sigma \mathcal{T}^{\mu \nu}\right)=\frac{1}{2}\left(\partial_\mu \xi_{1 \nu}+\partial_\nu \xi_{1 \mu}\right) \partial_\sigma\left(\xi_2^\sigma \mathcal{T}^{\mu \nu}\right)=\frac{1}{d}\left(\partial_\alpha \xi^\alpha\right) \partial_\sigma\left(\xi_2^\sigma T_\mu^\mu\right)=0~,
\eeq
using the conformal Killing equation and the tracelessness of the stress tensor. Combining the above we find the final result for the symplectic form to be:\footnote{Of course, in hindsight we could have expected this result. If we introduce the following functions
\beq
f_{\xi} \equiv \int_B d S_\mu \xi_\nu \mathcal{T}^{\mu \nu}
\eeq
on phase space, \eqref{eq:bulkcommutator} is simply the familiar statement that the symplectic form computes the (classical) Poisson bracket via
\beq
\omega\left(\xi_1,\xi_2\right)=\left\{f_{\xi_1}, f_{\xi_2}\right\}=-f_{\left[\xi_1, \xi_2\right]}~.
\eeq}
\beq 
\omega\left(\delta_{\zeta_1} g, \delta_{\zeta_2} g\right)=\frac{1}{2\pi}\int_A d S_\mu\left[\xi_1, \xi_2\right]_\nu \mathcal{T}^{\mu \nu}~.
\eeq
This result involving the commutator of vector fields indeed resembles the structure of the ($H_{\rm mod}$-component of the) Berry curvature \eqref{eq:Fmod}, when we evaluate it in the original state $|\Psi\rangle$.

\subsection{Stress tensor transformation law}\label{sec:Ttransform}
Let us for completeness derive the stress tensor transformation law that was used in \eqref{eq:stresstransf}. We start from the most general ansatz consistent with linearity of $\delta_{\xi} T$ and dimensional analysis:
\begin{align}
\delta_{\xi} \mathcal{T}^{\mu \nu} & =c_1 \xi^\sigma \partial_\sigma \mathcal{T}^{\mu \nu}+c_2\left(\partial_\sigma \xi^\sigma\right) \mathcal{T}^{\mu \nu}+c_3\left(\partial_\sigma \xi^\mu\right) \mathcal{T}^{\sigma \nu}+c_4\left(\partial^\nu \xi_\sigma\right) \mathcal{T}^{\mu \sigma} \nonumber \\
& \label{eq:transformationlawT} +c_5 \xi_\sigma \partial^\mu \mathcal{T}^{\sigma \nu}+c_6\left(\partial_\sigma \xi^\nu\right) \mathcal{T}^{\mu \sigma}+c_7\left(\partial^\mu \xi_\sigma\right) \mathcal{T}^{\sigma \nu}+c_8 \xi_\sigma \partial^\nu \mathcal{T}^{\mu \sigma}~.
\end{align}
We want to further constrain \eqref{eq:transformationlawT} using symmetry, tracelesness and conservation of $\delta_{\xi} \mathcal{T}^{\mu \nu}$. Recall that the conformal Killing equation is
\be \label{eq:Killingequation}
\partial_\mu \xi_\nu+\partial_\nu \xi_\mu=\frac{2}{d} \eta_{\mu \nu} \partial_\sigma \xi^\sigma~.
\ee
This can be used to write the terms multiplying the coefficients $c_6, c_7$ in terms of those involving $c_2, c_3$ and $c_4$. We have 
\begin{align} \label{eq:eq11}
& \left(\partial_\sigma \xi^\nu\right) \mathcal{T}^{\mu \sigma}=-\left(\partial^\nu \xi_\sigma\right) \mathcal{T}^{\mu \sigma}+\frac{2}{d}\left(\partial_\sigma \xi^\sigma\right) \mathcal{T}^{\mu \nu}~,  \\ 
\label{eq:eq12} & \left(\partial^\mu \xi_\sigma\right) \mathcal{T}^{\sigma \nu}=-\left(\partial_\sigma \xi^\mu\right) \mathcal{T}^{\sigma \nu}+\frac{2}{d}\left(\partial_\sigma \xi^\sigma\right) \mathcal{T}^{\mu \nu}~.
\end{align}
Therefore, these terms can be removed from the ansatz \eqref{eq:transformationlawT}. Moreover, imposing that $\delta_{\xi} \mathcal{T}^{\mu \nu}$ is symmetric under interchange of $\mu$ and $\nu$ gives
\be
c_3=-c_4~, \hspace{15pt} c_5=c_8~.
\ee
Indeed, for the first equation one can use that
\be
\left(\partial^\nu \xi_\sigma\right) \mathcal{T}^{\mu \sigma}-\left(\partial_\sigma \xi^\mu\right) \mathcal{T}^{\sigma \nu}=\left(\partial^\mu \xi_\sigma\right) \mathcal{T}^{\nu \sigma}-\left(\partial_\sigma \xi^\nu\right) \mathcal{T}^{\sigma \mu}~,
\ee
by subtracting \eqref{eq:eq11} from \eqref{eq:eq12}, and flipping $\mu$ with $\nu$.
In this way, we have reduced the ansatz to
\begin{align}
\delta_{\xi} \mathcal{T}^{\mu \nu} & =c_1 \xi^\sigma \partial_\sigma \mathcal{T}^{\mu \nu}+c_2\left(\partial_\sigma \xi^\sigma\right) \mathcal{T}^{\mu \nu}+c_3\left(\left(\partial_\sigma \xi^\mu\right) \mathcal{T}^{\sigma \nu}-\left(\partial^\nu \xi_\sigma\right) \mathcal{T}^{\mu \sigma}\right) \nonumber \\
& +c_5\left(\xi_\sigma \partial^\mu \mathcal{T}^{\sigma \nu}+\xi_\sigma \partial^\nu \mathcal{T}^{\mu \sigma}\right) .
\end{align}
Let us now impose conservation, using that $\partial_\mu \mathcal{T}^{\mu \sigma}=0$. We will need the following identity:
\be \label{eq:identity}
\partial_\mu \partial^\nu \xi_\sigma=\frac{1}{d}\left(\delta_\sigma^\nu \partial_\mu \partial_\alpha \xi^\alpha+\delta_{\sigma \mu} \partial^\nu \partial_\alpha \xi^\alpha-\delta_\mu^\nu \partial_\sigma \partial_\alpha \xi^\alpha\right)~.
\ee
This can be derived by acting with another derivative on the conformal Killing equation \eqref{eq:Killingequation}, and taking three different permutations:
\begin{align} \label{eq:eqcKeq1}
\partial_\mu \partial^\nu \xi_\sigma+\partial^\nu \partial_\sigma \xi_\mu & =\frac{2}{d}\left(\partial^\nu \partial_\alpha \xi^\alpha\right) \delta_{\mu \sigma}~,  \\
\label{eq:eqcKeq2} \partial_\mu \partial^\nu \xi_\sigma+\partial_\mu \partial_\sigma \xi^\nu & =\frac{2}{d}\left(\partial_\mu \partial_\alpha \xi^\alpha\right) \delta_\sigma^\nu~,  \\
\label{eq:eqcKeq3} \partial_\sigma \partial^\nu \xi_\mu+\partial_\mu \partial_\sigma \xi^\nu & =\frac{2}{d}\left(\partial_\sigma \partial_\alpha \xi^\alpha\right) \delta_\mu^\nu~.
\end{align}
Adding \eqref{eq:eqcKeq1} and \eqref{eq:eqcKeq2} and subtracting \eqref{eq:eqcKeq3} leads to \eqref{eq:identity}. In particular, \eqref{eq:identity} can be used to show that
\be \label{eq:identityforstress}
\partial_\mu \partial^\nu \xi_\sigma \mathcal{T}^{\mu \sigma}=0~.
\ee
We can use this equation to write
\begin{align}
\partial_\mu \delta_{\xi} &\mathcal{T}^{\mu \nu}
=\left(c_1+c_3\right)\left(\partial_\mu \xi^\sigma\right) \partial_\sigma \mathcal{T}^{\mu \nu}+\left(c_2+c_3\right)\left(\partial_\mu \partial_\sigma \xi^\sigma\right) \mathcal{T}^{\mu \nu} \nonumber \\
&\hspace{100pt}+c_5\left(\left(\partial_\mu \xi_\sigma\right) \partial^\mu \mathcal{T}^{\sigma \nu}+\xi_\sigma \partial_\mu \partial^\mu \mathcal{T}^{\sigma \nu}+\left(\partial_\mu \xi_\sigma\right) \partial^\nu \mathcal{T}^{\mu \sigma}\right)~.
\end{align}
From this we see that the coefficients are further constrained to satisfy
\be
c_1=-c_3=c_2~, \hspace{15pt} c_5=0 .
\ee
Hence, we find the transformation law
\be
\delta_{\xi} \mathcal{T}^{\mu \nu}=c_1\left(\xi^\sigma \partial_\sigma \mathcal{T}^{\mu \nu}+\left(\partial_\sigma \xi^\sigma\right) \mathcal{T}^{\mu \nu}-\left(\partial_\sigma \xi^\mu\right) \mathcal{T}^{\sigma \nu}+\left(\partial^\nu \xi_\sigma\right) \mathcal{T}^{\mu \sigma}\right)~,
\ee
which is automatically traceless. The overall normalization $c_1=1$ is fixed by requiring, for example, that a translation $\xi^\sigma=\delta_\sigma^\sigma$ acts via a derivative
\be
\delta_{\xi} \mathcal{T}^{\mu \nu}=\partial_\sigma \mathcal{T}^{\mu \nu}~.
\ee
This proves the stress tensor transformation law~\eqref{eq:stresstransf}:
\beq
\delta_{\xi} \mathcal{T}^{\mu \nu}=\xi^{\sigma} \partial_\sigma \mathcal{T}^{\mu \nu}+\left(\partial_\sigma \xi^\sigma\right) \mathcal{T}^{\mu \nu}-\left(\partial_\sigma \xi^\mu\right) \mathcal{T}^{\sigma \nu}+\left(\partial^\nu \xi_\sigma\right) \mathcal{T}^{\mu \sigma}~.
\eeq

\bibliographystyle{JHEP}
\bibliography{virasoro.bib}

\end{document}